\documentclass[reprint,amsmath,amssymb,aps,superscriptaddress]{revtex4-2}

\usepackage{graphicx}
\usepackage{hyperref}
\usepackage{siunitx}
\usepackage{dcolumn}
\usepackage{bm}
\usepackage{color}
\usepackage{xcolor}
\usepackage{bbm}

\makeatletter
\newcommand{\supplementarytableofcontents}{%
    \begingroup
    \setcounter{tocdepth}{1}%
    \begin{center}
        {\large\bfseries Contents}
    \end{center}
    \vspace{0.5em}%
    \@starttoc{snt}%
    \endgroup
}
\newcommand{\supplementarynote}[2]{%
    \section*{Supplementary Note #1. #2}%
    \addcontentsline{snt}{suppnote}{Supplementary Note #1. #2}%
}
\newcommand*\l@suppnote{\@dottedtocline{1}{0em}{0em}}
\makeatother

\newcommand{\ua}{\uparrow}
\newcommand{\da}{\downarrow}
\newcommand{\AB}{\mathrm{AB}}
\newcommand{\Vis}{\widetilde{\mathcal{V}}}
\DeclareMathOperator{\sgn}{sgn}
\newcommand{\mb}{\mathbf}
\renewcommand{\O}{\mathcal{O}}

\begin{document}
\title{Interferometry reveals spin-singlet fractional quantum Hall edges}

\author{R. Ayache} \email{equal contribution}
\affiliation{SPEC, CEA, CNRS, Université Paris-Saclay, CEA Saclay, 91191 Gif-sur-Yvette Cedex, France}
\author{K. Kim} \email{equal contribution}
\affiliation{Department of Physics, Korea Advanced Institute of Science and Technology, Daejeon 34141, Korea}
\author{M. Kuiri} \email{equal contribution}
\affiliation{SPEC, CEA, CNRS, Université Paris-Saclay, CEA Saclay, 91191 Gif-sur-Yvette Cedex, France}
\affiliation{Department of Physics, Birla Institute of Technology and Science, Pilani, Hyderabad Campus, Jawahar Nagar, Kapra Mandal, Medchal District, Telangana 500078, India}
\author{Q. Benichou}
\affiliation{SPEC, CEA, CNRS, Université Paris-Saclay, CEA Saclay, 91191 Gif-sur-Yvette Cedex, France}
\author{H. Chakraborti}
\affiliation{SPEC, CEA, CNRS, Université Paris-Saclay, CEA Saclay, 91191 Gif-sur-Yvette Cedex, France}
\author{L. Pugliese}
\affiliation{SPEC, CEA, CNRS, Université Paris-Saclay, CEA Saclay, 91191 Gif-sur-Yvette Cedex, France}
\author{K. Watanabe}
\affiliation{National Institute for Materials Science, 1-1 Namiki, Tsukuba 305-0044, Japan}
\author{T. Taniguchi}
\affiliation{National Institute for Materials Science, 1-1 Namiki, Tsukuba 305-0044, Japan}
\author{H.-S. Sim} \email{Corresponding author. hs\_sim@kaist.ac.kr}
\affiliation{Department of Physics, Korea Advanced Institute of Science and Technology, Daejeon 34141, Korea}
\author{P. Roulleau} \email{Corresponding author. preden.roulleau@cea.fr}
\affiliation{SPEC, CEA, CNRS, Université Paris-Saclay, CEA Saclay, 91191 Gif-sur-Yvette Cedex, France}




\begin{abstract}
	
The spin-unpolarized 2/3 fractional quantum Hall (FQH) state is an SU(2) multicomponent phase expected to host counterpropagating charge and neutral spin edge modes. Yet their properties, including spin polarization, coherence, spin-charge separation, and anyonic fusion of spinons, remain elusive because FQH edge measurements lack spin resolution. Here, we develop a spin-sensitive probe of graphene FQH edges using a pn junction between target FQH and probe integer states. An Aharonov--Bohm interferometer forms when their edge channels carry the same spin. Spin-dependent interference identifies unpolarized and polarized 2/3 edges at lower and higher magnetic fields, respectively. We observe multiparticle interference involving opposite-spin electrons, enabled by spin-charge separation and subsequent SU(2)$_1$ fusion of their spinons into the vacuum channel, gluing two otherwise independent single-electron interference processes into the correlated two-electron interference. Thus edge transport reveals SU(2) spin-singlet topological order, with implications for parafermions.

\end{abstract}

\maketitle

\vspace{1em}

Multicomponent FQH systems~\cite{Halperin1983,Wen1992,Girvin1996} host distinct correlated phases involving spin, valley, layer, or subband degrees of freedom. These phases compete with single-component states at a given filling factor $\nu$, leading to phase transitions as external fields vary. Probing these internal degrees of freedom can thus reveal the nature of competing phases and their fractionalization. Here, we demonstrate experimental access to the spin of edge channels at $\nu=2/3$ in graphene.

The $\nu=2/3$ system provides a paradigmatic setting for studying spin and charge in multicomponent FQH effects. Its competing phases are the spin-singlet FQH state---the singlet ground state of the SU(2)-symmetric Coulomb interaction~\cite{Maksym1989,Eisenstein1990,Wu1993,Kraus2002,Stern2004,Zhang1984}---and the fully spin-polarized state, the particle-hole conjugate of the single-component $\nu=1/3$ Laughlin state~\cite{Johnson1991}. The latter is favored at higher magnetic fields by the larger Zeeman energy. 

These phases support distinct edge structures~\cite{Balatsky1991,Moore1997,Imura1998,Wu2012,Johnson1991,Kane1994,Bid2010,Lafont2019,Wang2021,Ponomarenko2024}.
The spin-polarized edge consists of a downstream $\nu=1$ channel and an upstream $\nu=1/3$-like channel, which reconstruct into downstream charge and upstream charge-neutral modes in the presence of strong disorder~\cite{Johnson1991,Kane1994}. By contrast, the spin-singlet state hosts a spin-unpolarized edge comprising a downstream charge mode and an upstream charge-neutral spin mode~\cite{Balatsky1991,Moore1997,Imura1998,Wu2012,Ponomarenko2024}. These modes have been predicted to enable nontrivial anyonic fractionalization and fusion. For instance, an electron injected onto this edge undergoes spin-charge separation~\cite{Auslaender2005,Jompol2009} into a chargon and a spinon. Spinons are Abelian anyons carrying spin $1/2$, described by the SU(2)$_1$ conformal field theory~\cite{Balatsky1991,Moore1997}. They obey the fusion rule~\cite{Gepner1986}
\begin{equation} \label{spinonfusion}
	\frac{1}{2}\times\frac{1}{2}=0,
\end{equation}
meaning that two spinons (labeled by $1/2$) fuse into the vacuum channel ($0$) as they approach each other. In particular, the fusion of an up-spinon and a down-spinon (spinons with spin up and down, respectively) is dominated by the identity operator $\mathbbm 1$, which corresponds to a spin singlet.
  
While the bulk properties~\cite{Eisenstein1990,Kraus2002,Stern2004} of the two competing phases and the upstream neutral mode of the polarized edge~\cite{Bid2010} have been experimentally established, direct characterization of the spin-unpolarized edge remains challenging. A previous experiment investigated this edge~\cite{Lafont2019}, but was primarily limited to charge transport. 
Its spin properties thus remain inaccessible to conventional charge and heat probes.
 
We develop an approach to identify the spin structure of FQH edge channels in graphene solely through edge transport, without external spin-sensitive spectroscopy. Our approach uses Aharonov--Bohm (AB) interferometers formed along a gate-defined bipolar junction between a target FQH edge of unknown spin structure and a probe integer quantum Hall (QH) edge of known spin polarization. Because interference occurs only when the target and probe edge channels carry the same spin, the interference directly probes the spin properties of the target edge. 
The presence of interference identifies the spin of the interfering channels through its AB period, distinguishing spin-unpolarized from polarized edges, while its visibility and winding numbers reveal spin coherence and fractionalization. In particular, we observe an AB interference of two opposite-spin electrons on the unpolarized edge, enabled by their spin-charge separation and the subsequent fusion of their spinons following Eq.~\eqref{spinonfusion}.

In graphene, the spin-unpolarized and polarized $\nu=2/3$ states are embedded in the fourfold spin--valley degeneracy of a Landau level. These states can be represented, respectively, by the filling factors $(1,1,1/3,1/3)$ and $(1,1,2/3,0)$ of the four spin-valley components~\cite{Sodemann2014,Hegde2022}.
In both states, the two fully filled components form the $\nu=0$ background. In the spin-unpolarized state, the remaining fractional filling is shared equally between components of opposite spin, whereas in the polarized state it occupies a single spin component.

\begin{figure*}
	\centering
	\includegraphics[width=1\textwidth]{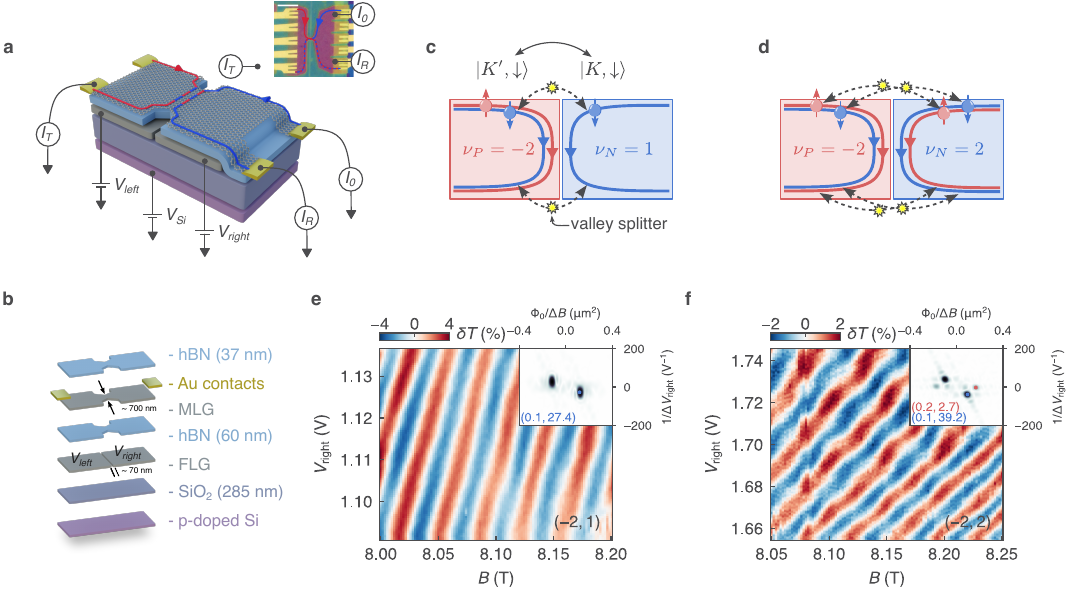}
	\caption{\textbf{Interferometric spin access in the integer QH regime.}
		\textbf{a,b,} Schematic of the device. Monolayer graphene is encapsulated in hBN and placed on a patterned graphite gate split into two electrically isolated halves. Gate voltages $V_{\mathrm{left}}$ and $V_{\mathrm{right}}$ independently tune the filling factors $\nu_{\mathrm{P}}$ and $\nu_{\mathrm{N}}$ in the left and right graphene regions, respectively. The graphite gates are separated by a \SI{70}{nm} slit, while the carrier density is further fine-tuned by the doped Si/SiO$_2$ (\SI{285}{nm}) back gate. The graphene is additionally etched above the slit, forming a \SI{700}{nm}-wide constriction.
		\textbf{c,d,} Edge-channel configurations for $(\nu_{\mathrm{P}},\nu_{\mathrm{N}})=(-2,1)$ in \textbf{c} and $(-2,2)$ in \textbf{d}.
		\textbf{e,f,} AB oscillations of the differential transmission $\delta \mathcal{T}$, measured at $(\nu_{\mathrm{P}},\nu_{\mathrm{N}})=(-2,1)$ in \textbf{e} and $(-2,2)$ in \textbf{f}, as a function of the applied magnetic field $B$ and the right-gate voltage $V_{\mathrm{right}}$. Insets: Fast Fourier transforms (FFTs) of the corresponding AB oscillations. The dominant FFT frequencies show that one spin-down Mach--Zehnder interferometer (MZI) is formed at $(\nu_{\mathrm{P}},\nu_{\mathrm{N}})=(-2,1)$, whereas both spin-down and spin-up MZIs coexist at $(-2,2)$.}
	\label{fig1}
\end{figure*}

\subsection*{Operating principle of interferometric spin access}

The device is shown in Fig.~\ref{fig1}a,b (Supplementary Notes 1-3). A monolayer graphene sheet is tuned into a bipolar QH p-n junction~\cite{Williams2007,Abanin2007}, with QH states at filling factors $\nu_{\mathrm{P}}$ and $\nu_{\mathrm{N}}$ on the left and right halves, respectively. Electron Mach--Zehnder interferometers (MZIs) have been demonstrated to form along the p-n interface for certain combinations of integer filling factors~\cite{Wei2017,Jo2021,Jo2022,Chakraborti2025}. We first illustrate the formation of the MZIs and how the spin of the edge channels can be inferred from the interference signal in the well-understood integer QH regime.

At $(\nu_{\mathrm{P}},\nu_{\mathrm{N}})=(-2,1)$, the p region hosts two clockwise edge channels of opposite spin, while the n region has a single counterclockwise spin-down edge channel. Along the p-n interface, all channels copropagate (Fig.~\ref{fig1}c). Here, spin-down denotes the Zeeman-favored spin orientation. At the upper (lower) intersection between the physical edge and the p-n interface, the incoming spin-down edge (interface) channel is scattered into the outgoing interface (edge) channel (dashed arrows in Fig.~\ref{fig1}c). Spin is conserved during this scattering, while the valley degree of freedom changes~\cite{Tworzydlo2007,Trifunovic2019}, resulting in a spin-down MZI~\cite{Wei2017,Jo2021,Jo2022,Chakraborti2025}. The two spin-down interface channels form the MZI arms, and the two intersections serve as the beam splitters. 

The transmission probability of the MZI is  $\mathcal{T}_{\mathrm{MZI}}=\mathcal{R}_1\mathcal{T}_2+\mathcal{T}_1\mathcal{R}_2+2\sqrt{\mathcal{R}_1\mathcal{T}_1\mathcal{R}_2\mathcal{T}_2}
\cos\!\left(\varphi_{\mathrm{AB}}+\phi\right)$, where $\mathcal{T}_1$ and $\mathcal{R}_1=1-\mathcal{T}_1$ denote the transmission and reflection probabilities at the upper beam splitter, respectively, and $\mathcal{T}_2$ and $\mathcal{R}_2=1-\mathcal{T}_2$ those at the lower beam splitter. Here $\varphi_{\mathrm{AB}}=2\pi BA/\Phi_0$ is the AB phase associated with the magnetic flux enclosed by the interferometer of area $A$, $\phi$ accounts for the scattering phases acquired at the beam splitters, and $\Phi_0=h/e$.

The enclosed area is tuned by changing the gate voltages $V_{\mathrm{left}}$ and $V_{\mathrm{right}}$, which shift the interface channels electrostatically~\cite{Flor2022}. An additional back-gate voltage $V_{\mathrm{Si}}$ further fine-tunes the separation between the interfering edge channels. Figure~\ref{fig1}e shows oscillations of the differential transmission $\delta \mathcal{T}$ as a function of $B$ and $V_{\mathrm{right}}$. The corresponding  fast Fourier transform (FFT) exhibits a single frequency, $\left(f_{B/\Phi_0},f_{V_{\mathrm{right}}}\right)=\left(\Phi_0/\Delta B,\,1/\Delta V_{\mathrm{right}}\right)=\left(0.12~\mu\mathrm{m}^2,-27.4~\mathrm{V}^{-1}\right)$. Given the \SI{700}{nm} arm length set by electron-beam lithography, we extract a separation of approximately \SI{170}{nm} between the two arms. The single frequency suggests the absence of disorder-induced tunneling between the interface channels, consistent with the smooth electrostatic potential defining the p-n junction.

\begin{figure*}
	\centering
	\includegraphics[width=1\textwidth]{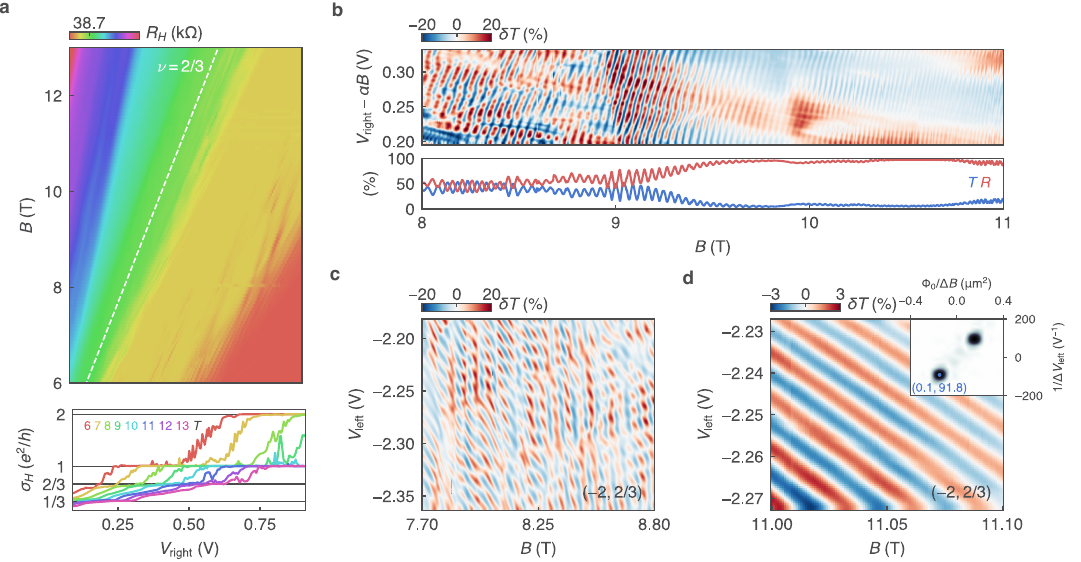}
	\caption{\textbf{Spin transition of the $2/3$ edge.}
		\textbf{a,} Upper panel: Two-point Hall resistance $R_{\mathrm{H}}$ measured on the right side of the graphene device as a function of $B$ and $V_{\mathrm{right}}$. The $\nu_{\mathrm{N}}=2/3$ state becomes well developed for $B > \SI{6}{T}$ and is visible as a green region. Lower panel: Corresponding Hall conductance $\sigma_{\mathrm{H}}$ measured at selected magnetic fields.
		\textbf{b,} Upper panel: Differential transmission $\delta \mathcal{T}$ through the p-n junction at $(\nu_{\mathrm{P}},\nu_{\mathrm{N}})=(-2,2/3)$ as a function of $B$ and the compensated gate voltage $V_{\mathrm{right}}-\alpha B$, with $\alpha \approx 0.124~\mathrm{V/T}$ accounting for the magnetic-field-induced shift of the $2/3$ plateau. A sharp change occurs near $B\approx \SI{9.5}{T}$: The interference visibility is reduced and the pattern evolves from multiple frequencies to a single frequency, indicating a sudden change in the interface-channel structure. This is consistent with the transition between the spin-unpolarized and spin-polarized $2/3$ states. Lower panel: Raw transmission $\mathcal{T}$ and reflection $\mathcal{R}$ traces measured at $V_{\mathrm{right}}-\alpha B=\SI{0.3}{V}$.
		\textbf{c,} AB oscillations in the $(-2,2/3)$ regime as a function of $B$ and $V_{\mathrm{left}}$ at low magnetic fields ($B\approx\SI{8}{T}$), where the spin-unpolarized $2/3$ state is expected. The interference pattern displays multiple frequencies.
		\textbf{d,} AB oscillations in the same $(-2,2/3)$ regime at higher magnetic fields ($B\approx\SI{11}{T}$), where the spin-polarized $2/3$ state is expected. Inset: FFT of the oscillations, showing a single frequency, $\left(f_{B/\Phi_0},f_{V_{\mathrm{left}}}\right)=(\SI{0.148}{\micro\meter^2},\,\SI{91.8}{V^{-1}})$.}
	\label{fig2}
\end{figure*}

At $(\nu_{\mathrm{P}},\nu_{\mathrm{N}})=(-2,2)$, the n region has an additional spin-up edge channel (Fig.~\ref{fig1}d). The AB oscillations now exhibit two distinct frequencies in the FFT (Fig.~\ref{fig1}f), demonstrating the coexistence of two MZIs: one formed by spin-down electrons, as in the $(-2,1)$ configuration, and the other by spin-up electrons. The total transmission is well described by $\mathcal{T}_{\mathrm{MZI,tot}}=\mathcal{T}_{\mathrm{MZI},1}(A_1) + \mathcal{T}_{\mathrm{MZI},2}(A_2)$, where $\mathcal{T}_{\mathrm{MZI},i}(A_i)$ denotes the transmission of the $i$-th interferometer enclosing area $A_i$. No higher-order frequencies appear in the FFT, consistent with two independently operating MZIs. Simulations of the interference patterns in the $(-2,1)$ and $(-2,2)$ regimes are shown in Extended Data Fig.~\ref{simu}. The extracted frequencies correspond to arm separations of approximately \SI{140}{nm} and \SI{240}{nm}. The difference in arm separation between the two MZIs arises from the electrostatic positioning of the edge channels~\cite{Flor2022}.

By contrast, no AB oscillation is observed at $(\nu_{\mathrm{P}},\nu_{\mathrm{N}})=(-1,1)$ (Extended Data Fig.~\ref{figED-1_1}). In this regime, the p and n regions each support a single interface channel with opposite spin, preventing the formation of an MZI.

The presence or absence of AB oscillations, together with their frequencies, suggests that the p-n interface can serve as a platform for detecting the spin of QH edge channels. If the spin polarization of a ``target" QH state in the n region is unknown whereas that of a ``probe" state in the p region is known, the presence of an AB oscillation implies that an interface channel of the target state carries the same spin as the probe channel, allowing an MZI to form. For example, the absence of oscillations at $(\nu_{\mathrm{P}},\nu_{\mathrm{N}})=(-1,1)$ indicates that the interface channel of the target state is spin-down, as spin-up probe electrons cannot couple to it. The appearance of a single oscillation frequency at $(-2,1)$ indicates that the additional probe channel has the matching spin-down polarization. The two frequencies at $(-2,2)$ demonstrate that the interface channels of the target state host both spin-up and spin-down electrons.
These observations agree with the expected configuration of p-n interface channels. 

\subsection*{Spin polarization transition of the 2/3 edge states}

We investigate the $\nu=2/3$ regime with the junction set to $(\nu_{\mathrm{P}},\nu_{\mathrm{N}})=(-2,2/3)$. Figure~\ref{fig2}a presents the Landau fan measured on the right half of the device, revealing an FQH state at $\nu_{\mathrm{N}}=2/3$ with a quantized Hall conductance of $\sigma_{\mathrm{H}}=2e^2/3h$. The junction transmission exhibits AB oscillations throughout the $\nu_{\mathrm{N}}=2/3$ plateau (Fig.~\ref{fig2}b). Around $B\approx\SI{9.5}{T}$, we observe a sharp drop in the transmission, indicative of a transition of the interface channel structure. This is consistent with the expected transition between the spin-unpolarized and spin-polarized $\nu=2/3$ states~\cite{Eisenstein1990,Kraus2002,Stern2004,Sodemann2014,Hegde2022}.

The observed AB oscillations demonstrate the formation of an interferometer, as in the integer QH regime in Fig.~\ref{fig1}. In contrast to the integer case, however, the interferometer comprises integer and fractional interface channels associated with the $\nu_{\mathrm{P}}=-2$ and $\nu_{\mathrm{N}}=2/3$ regions, respectively. Although one arm is fractional, scattering at the beam splitters transfers electrons rather than fractional charges between the two channels. Consequently, the AB period is set by the electron charge. 

To identify the nature of the $\nu=2/3$ state, we first measure the AB oscillations around $B\approx\SI{8}{T}$, where the state is expected to be spin unpolarized (Fig.~\ref{fig2}c). In this regime, the oscillations display a complex interference pattern with multiple frequency components, suggesting an intricate edge-channel structure of the spin-unpolarized state. A detailed analysis presented later provides further insight into this edge structure.

We next perform the same measurement at higher magnetic fields, around $B\approx \SI{11}{T}$ with $(\nu_{\mathrm{P}},\nu_{\mathrm{N}})=(-2,2/3)$. In contrast to the low-field regime, the AB oscillations exhibit a regular pattern with a single frequency (Fig.~\ref{fig2}d),
consistent with a single interferometer.
To further determine the spin of the interfering channel, we then measure $(\nu_{\mathrm{P}},\nu_{\mathrm{N}})=(-1,2/3)$, thereby removing the spin-down channel from the p region. No AB oscillations are observed (Extended Data Fig.~\ref{figED-1_23}), in contrast to the $(-2,2/3)$ configuration. This shows that the interface channel of the $\nu_{\mathrm{N}}=2/3$ state in this high-field regime is spin polarized in the spin-down direction---the direction favored by the Zeeman energy.

\subsection*{Spin-unpolarized edge at $\nu_{\mathrm{N}}=2/3$}

\begin{figure*}
	\centering
	\includegraphics[width=1\textwidth]{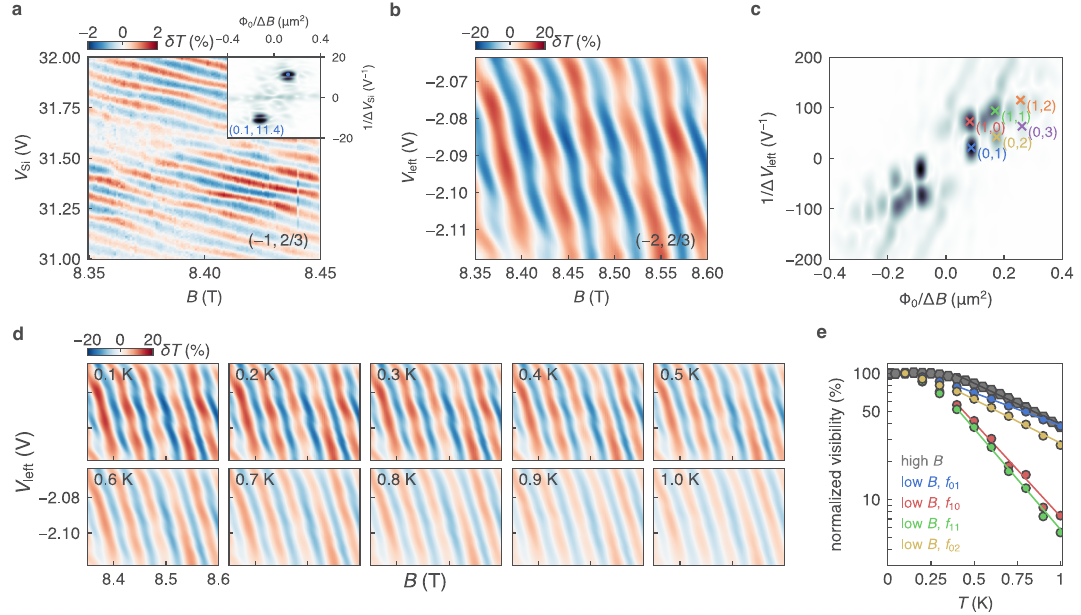}
	\caption{\textbf{Interference signatures of the spin-unpolarized $\nu_{\mathrm{N}}=2/3$ edge.}
		\textbf{a,} AB oscillations in the $(\nu_{\mathrm{P}},\nu_{\mathrm{N}})=(-1,2/3)$ regime at low magnetic fields, as a function of $B$ and silicon-gate voltage $V_{\mathrm{Si}}$. Inset: FFT of the oscillations showing a single frequency
		$(\Phi_0/\Delta B,\,1/\Delta V_{\mathrm{Si}})
		=(\SI{0.121}{\micro\meter^2},\,\SI{11.4}{V^{-1}})$. 
		\textbf{b,} AB oscillations at $(\nu_{\mathrm{P}},\nu_{\mathrm{N}})=(-2,2/3)$.
		\textbf{c,} FFT of the oscillations in \textbf{b}, showing multiple frequencies that are linear combinations of two frequencies	$f_{10}
		=(\SI{0.082}{\micro\meter^2},\,\SI{72.8}{V^{-1}})$ and
		$f_{01}=\left(f_{B/\Phi_0},f_{V_{\mathrm{left}}}\right)
		=(\SI{0.087}{\micro\meter^2},\,\SI{21.2}{V^{-1}})$,
		labelled $(1,0)$ and $(0,1)$, respectively.		
		 Higher-order frequencies, labelled $(n,m)$, satisfy
		$f_{nm}=nf_{10}+mf_{01}$.
		\textbf{d,} Temperature dependence of the AB oscillations under the same conditions as in \textbf{b}.
        \textbf{e,} Temperature dependence of the interference visibilities of the $f_{01}$, $f_{10}$, $f_{11}$, and $f_{02}$ components in the spin-unpolarized regime, shown in blue, red, green, and yellow, respectively, and extracted from \textbf{d}. The visibility measured in the spin-polarized regime is shown in grey for comparison.
        Each visibility is defined as the amplitude of the corresponding frequency component of the zero-bias conductance, normalized to its value at $\SI{100}{mK}$.
		}
	\label{fig3}
\end{figure*}

We investigate the low-field regime around $B=\SI{8.4}{T}$, first measuring the junction at $(\nu_{\mathrm{P}},\nu_{\mathrm{N}})=(-1,2/3)$. The AB oscillations exhibit a single frequency (Fig.~\ref{fig3}a), demonstrating that the fractional interface channel of the $\nu_{\mathrm{N}}=2/3$ state can couple to spin-up probe electrons. We then set $\nu_{\mathrm{P}}=-2$, thereby introducing an additional spin-down probe channel. The AB oscillations now exhibit multiple frequencies $f_{nm}$ (Figs.~\ref{fig3}b,c). This shows that spin-down electrons also participate in the AB oscillations. Together, these measurements establish that the interface channel of the $\nu_{\mathrm{N}}=2/3$ state is spin unpolarized below $B\approx\SI{9.5}{T}$, in contrast to the spin-polarized state around $B \approx \SI{11}{T}$ in Fig.~\ref{fig2}d.

The two dominant frequencies, denoted by $f_{10}$ and $f_{01}$, arise from single windings around their respective AB loops. The different AB areas identify $f_{10}$ and $f_{01}$ with spin-up and spin-down electrons, respectively, consistent with the spin-up channel in the p region lying closer to the n-region channel than the spin-down channel.

Others satisfy $f_{nm}=n f_{10}+m f_{01}$ with $n,m=0,1,2,\ldots$, establishing $f_{10}$ and $f_{01}$ as the fundamental frequencies. 
These higher-order frequencies arise from interference involving $n$ windings of spin-up electrons and $m$ windings of spin-down electrons around the AB loops. 
These multiple-winding processes result from interactions mediated by the fractional interface channel in the n region, as only a single winding is allowed in the noninteracting MZI of the integer QH regime (Fig.~\ref{fig1}). In particular, $f_{11}$ represents interference involving two electrons with opposite spins. This differs from known phenomena involving two noninteracting, indistinguishable electrons, such as the Hanbury Brown and Twiss effect~\cite{Neder2007} and Hong--Ou--Mandel effect~\cite{Chakraborti2025,Dubois2013,Bocquillon2013}. The presence of $f_{11}$ therefore reflects nontrivial interaction effects, as discussed below.

\subsection*{Two-electron interference via spinon fusion}

\begin{figure*}
	\centering
	\includegraphics[width=1\textwidth]{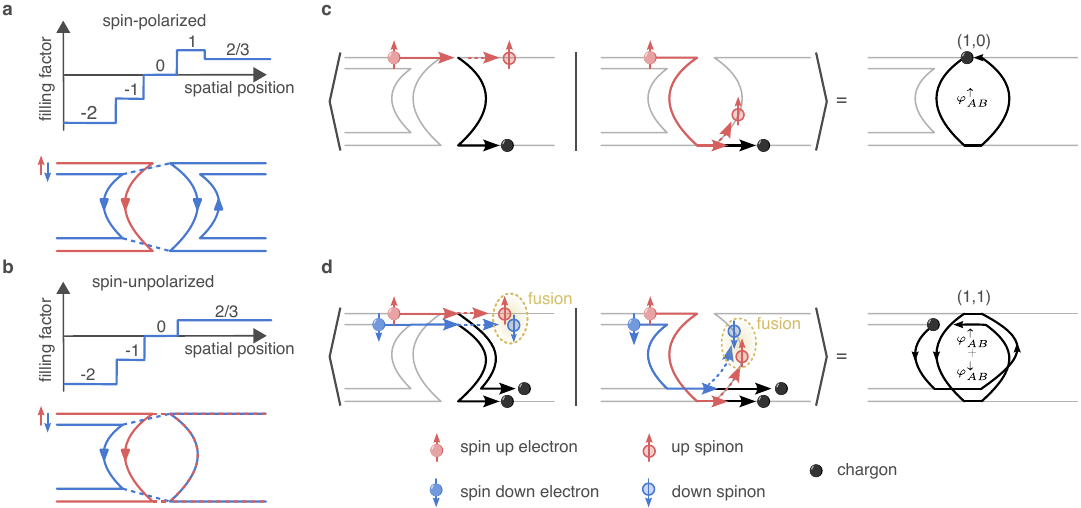}
	\caption{\textbf{Theoretical model.}
		\textbf{a,b,} Spatial variation of the filling factor across the p-n interface (upper) and corresponding interface channels (lower panels) in the spin-polarized and unpolarized regimes for $(\nu_{\mathrm{P}},\nu_{\mathrm{N}})=(-2,2/3)$. 
		Spin-up and spin-down channels are shown in red and blue, respectively, while the spin-unpolarized fractional channel is shown as alternating red and blue dashed lines. Electron tunneling between channels is depicted by dotted lines. 
		\textbf{c,} An interference process contributing to $f_{10}$ in the spin-unpolarized regime. The first two panels depict the interfering subprocesses in schematic bra-ket notation, and the third depicts their overlap involving a single winding. In each subprocess, a spin-up electron fractionalizes into a downstream chargon (black dot) and an upstream up-spinon (red up-arrow). The chargons generated in the two subprocesses overlap well, whereas the oppositely propagating spinons remain separated by a time delay $\Delta t_s$ and thus overlap only weakly.
		\textbf{d,} An interference process contributing to $f_{11}$. In one subprocess (left panel), both a spin-up electron and a spin-down electron enter the fractional channel through the upper beam splitter, whereas in the other subprocess (middle panel), both enter through the lower beam splitter. The electrons fractionalize into chargons and spinons, and the resulting up- and down-spinons fuse into the identity operator $\mathbbm 1$ of the vacuum channel in each subprocess [see Methods Eq.~\eqref{spinonOPE}]. The fusion glues together two otherwise independent single-electron interference processes into two-electron interference in which one charge winds around each of the spin-up and spin-down AB loops.  
	}
	\label{fig4}
\end{figure*}

The experimental observations are consistent with theoretical expectations. Figure~\ref{fig4}a shows the expected interface channels at $(\nu_{\mathrm{P}},\nu_{\mathrm{N}})=(-2,2/3)$ in the spin-polarized regime. In the absence of disorder, the $\nu_{\mathrm{N}}=2/3$ state supports an upstream charge channel between the $\nu=2/3$ and $\nu=1$ regions, and a downstream charge channel between $\nu=1$ and $\nu=0$~\cite{Johnson1991}. The AB oscillations in Fig.~\ref{fig2}d are described by an MZI formed between the spin-down downstream channels in the p and n regions (Supplementary Note~4). 
The single-frequency oscillations suggest that the interface is free of disorder that induces electron tunneling between the two counterpropagating channels of $\nu_{\mathrm{N}}=2/3$, in contrast to disorder-driven edge reconstruction~\cite{Kane1994}. The absence of additional frequencies further indicates strongly suppressed tunneling between the spin-down probe channel and the inner upstream channel of the fractional edge, consistent with their spatial separation due to the electrostatic confinement.

Figure~\ref{fig4}b shows the expected interface channels at $(\nu_{\mathrm{P}},\nu_{\mathrm{N}})=(-2,2/3)$ in the spin-unpolarized regime. In the n region, an interface channel forms between the $\nu_{\mathrm{N}}=2/3$ and $\nu=0$ regions. This channel is predicted~\cite{Balatsky1991,Moore1997,Imura1998,Wu2012} to host a downstream charge mode and an upstream charge-neutral spin mode, hallmarks of spin-charge separation, when SU(2) symmetry is preserved. The resulting interferometer is not a conventional single-spin MZI but combines the two opposite-spin integer channels in the p
region with the fractional channel in the n region, giving rise to
AB oscillations at multiple frequencies, $f_{nm}=n f_{10}+m f_{01}$, as observed in Fig.~\ref{fig3}c.  

Figure~\ref{fig4}c shows an $f_{10}$ process, in which a spin-up electron
enters the fractional channel from the p region via the upper or lower beam splitter in the two subprocesses (Fig.~\ref{fig4}c, left and middle).
Upon entering, the electron undergoes spin-charge separation into a
downstream charge excitation (chargon) and an upstream spin
excitation (up-spinon). The interference involves a single charge winding around the AB loop (Supplementary Note~5). The $f_{01}$ interference occurs analogously.

Figure~\ref{fig4}d illustrates an $f_{11}$ process, involving both a spin-up electron and a spin-down electron. They enter the fractional channel from the p region, either through the upper or the lower beam splitter (Fig.~\ref{fig4}d, left and middle panels). Upon entering, they undergo spin-charge separation, producing two chargons, an up-spinon, and a down-spinon. The spinons are spin-1/2 anyonic excitations~\cite{Balatsky1991,Moore1997}, and fuse into the vacuum channel when they are close to each other, following the SU(2)$_1$ fusion rule in Eq.~\eqref{spinonfusion}. The chargons generated at the upper beam splitter in one subprocess propagate downstream and overlap with those generated at the lower beam splitter in the other subprocess. The spinon fusion glues together the AB trajectories of the spin-up and spin-down electrons, giving rise to an $f_{11}$ interference involving one charge winding around each of the spin-dependent AB loops. The gluing corresponds to either a spin swap or no swap from the perspective of the two electrons. Our theory confirms the presence of these processes (Supplementary Note~6), accounting for the observation in Fig.~\ref{fig3}c.

Interestingly, the two-electron $f_{11}$ interference is more robust against thermal decoherence than expected from a naive factorization into the two single-electron interferences, $f_{10}$ and $f_{01}$. Figure~\ref{fig3}e shows that the measured visibility $\mathcal{V}_{11}$  of the $f_{11}$ interference decays more slowly than the product $\mathcal{V}_{10} \mathcal{V}_{01}$ of the $f_{10}$ and $f_{01}$ visibilities as temperature $T$ increases,
\begin{equation} \label{thermalvisibilities}
\frac{\mathcal{V}_{11}}{\mathcal{V}_{10} \mathcal{V}_{01} }\propto e^{ 2 \zeta T } \quad \quad \text{with} \quad \zeta > 0.
\end{equation}
This originates from the spinon fusion, as shown below.

We first consider the single-electron $f_{10}$ interference (Fig.~\ref{fig4}c). It requires spatial overlap between corresponding excitations generated in the two interfering subprocesses. The chargon generated at the upper beam splitter in one subprocess travels downstream with velocity $v_{\mathrm{N}c}$ and reaches the lower beam splitter in time $l/v_{\mathrm{N}c}$, where it overlaps with the chargon generated there in the other subprocess (Methods). Here, $l$ is the interferometer arm length. The spinon, however, travels upstream with velocity $v_{\mathrm{N}s}$ during this time, away from the lower beam splitter. The spinons generated in the two subprocesses are therefore separated by a time delay $\Delta t_s=l/v_{\mathrm{N}s}+l/v_{\mathrm{N}c}$. This delay suppresses the interference visibility $\mathcal{V}_{10}$~\cite{LeHur2005,Kim2009} by a factor $e^{-\zeta T}$, with $\zeta = \pi k_{\mathrm B} \Delta t_s/(2\hbar)$ (Methods). The same factor suppresses the visibility $\mathcal{V}_{01}$ of the $f_{01}$ interference. This suppression is absent in the spin-polarized regime, where spin-charge separation does not happen, consistent with the experimental observation (Fig.~\ref{fig3}e) of the slower thermal decay in the spin-polarized regime.

By contrast, in the two-electron $f_{11}$ interference (Fig.~\ref{fig4}d), the up- and down-spinons fuse into the vacuum channel, leaving no spinon to propagate upstream. Consequently, its visibility $\mathcal{V}_{11}$ is free from the spinon-induced thermal decoherence ($e^{-\zeta T}$) that suppresses both single-electron visibilities $\mathcal{V}_{10}$ and $\mathcal{V}_{01}$ (Methods).	
This explains Eq.~\eqref{thermalvisibilities}. Comparing the experimental data in Fig.~\ref{fig3}e  with Eq.~\eqref{thermalvisibilities} gives $\zeta=\SI{0.45}{K^{-1}}$, corresponding to a spinon delay time $\Delta t_s$ of a few picoseconds (Methods). This is of the same order of magnitude as $l/v$ for the arm length $l=\SI{700}{nm}$ and a typical edge velocity of $v\sim 10^5$ m/s.

The frequencies $f_{nm}$ persist despite weak breaking of the SU(2)
symmetry of the fractional channel (Supplementary Note~8).

\section*{Perspective}

The spin-unpolarized fractional channel of the spin-singlet $\nu=2/3$ state,
long experimentally unexplored, is directly identified through interferometric spin access. 
The observed AB interference with multiple windings reveals many-body coherent dynamics in the
channel. In particular, the $f_{11}$ frequency reveals a new type of multiparticle interference involving two
distinguishable electrons with opposite spins and the SU(2)$_1$ anyonic fusion of opposite spinons into the vacuum channel.

The fractional channel thus provides a unique platform for realizing a one-dimensional chiral metal with zero net spin under a strong magnetic field and for investigating fractionalizations, fusions, and anyonic quasiparticles~\cite{Nakamura2020,Bartolomei2020,Lee2023}.

Our observation of the spin-unpolarized fractional channel in graphene opens opportunities to explore the coupling between fractionalization and Cooper pairing, using superconducting contacts~\cite{Gul2022,Vignaud2023}. One direction is to realize parafermionic zero modes~\cite{Clarke2014,Mong2014,Wu2018}. Our identification of the
spin-unpolarized edge channel and its quantum coherence provides key prerequisites for this direction. Another is to investigate the spin-resolved interplay between Andreev reflection and spin-charge separation in superconductor-FQH hybrids.

High-quality, gate-defined p-n junctions provide a versatile platform for probing spin structures, coherent dynamics, and correlations in edge channels across FQH states in van der Waals or Dirac materials. Combining the p-n junction with spin-selective injection or detection or shot noise measurements can reveal the interplay among topology, interactions, and multiple edge components, opening avenues for new functionalities.

\section*{METHODS}

\subsection*{Model for the spin-unpolarized regime}

We model the interferometer formed along the
$(\nu_{\mathrm{P}},\nu_{\mathrm{N}})=(-2,2/3)$ p-n interface in the spin-unpolarized
regime in Fig.~\ref{fig4}b. 
We consider the SU(2)-symmetric $\nu_{\mathrm{N}}=2/3$ state. The channels forming
the interferometer arms are described by the Hamiltonian
\begin{equation}
\begin{aligned}
H_0 ={}&
-\sum_{\sigma=\uparrow,\downarrow}
\frac{i\hbar v_{\mathrm{P}\sigma}}{2\pi}
\int dx\,
\psi_{\mathrm{P}\sigma}^{\dagger}
\partial_x
\psi_{\mathrm{P}\sigma}
\\
&+
\sum_{\alpha=c,s}
\frac{\hbar v_{\mathrm{N}\alpha}}{4\pi}
\int dx\,
\left(\partial_x\phi_{\mathrm{N}\alpha}\right)^2 .
\end{aligned}
\label{eq:H0}
\end{equation}
Here $\psi_{\mathrm{P}\sigma}$ describes spin-$\sigma$ electrons on the
$\nu_{\mathrm{P}}=-2$ region ($\sigma=\uparrow,\downarrow$), while
$\phi_{\mathrm{N}c}$ and $\phi_{\mathrm{N}s}$ describe the charge and spin modes in the
$\nu_{\mathrm{N}}=2/3$ region. Their velocities are denoted by $v_{\mathrm{P}\sigma}$, $v_{\mathrm{N}c}$, and $v_{\mathrm{N}s}$. The bosonic fields satisfy $[\phi_{\mathrm{N}c}(x_1),\phi_{\mathrm{N}c}(x_2)] = i\pi\,\mathrm{sgn}(x_1-x_2)$, $[\phi_{\mathrm{N}s}(x_1),\phi_{\mathrm{N}s}(x_2)] = -i\pi\,\mathrm{sgn}(x_1-x_2)$, and $[\phi_{\mathrm{N}c}(x_1),\phi_{\mathrm{N}s}(x_2)] = 0$.
The spin-up and spin-down electron operators on the
$\nu_{\mathrm{N}}=2/3$ channel are $\psi_{\mathrm{N}\uparrow}^{\dagger} = \frac{F_{\mathrm{N}\uparrow}^{\dagger}}{a} e^{i\sqrt{3/2} \phi_{\mathrm{N}c}} e^{i\sqrt{1/2}\phi_{\mathrm{N}s}}$ and $\psi_{\mathrm{N}\downarrow}^{\dagger} = \frac{F_{\mathrm{N}\downarrow}^{\dagger}}{a} e^{i\sqrt{3/2} \phi_{\mathrm{N}c}} e^{-i\sqrt{1/2}\phi_{\mathrm{N}s}}$.
Here $F_{\mathrm{N}\sigma}^{\dagger}$ are Klein factors satisfying $\{F_{\mathrm{N}\uparrow},F_{\mathrm{N}\downarrow}\} =\{F_{\mathrm{N}\uparrow},F_{\mathrm{N}\downarrow}^{\dagger}\} =0$ and $a$ is a short-distance cutoff. 
For simplicity, we assume that the interface channels extend to the physical edges; this does not affect the interference mechanisms.

We next consider the upper and lower beam splitters at $x=0$ and $x=l$, where $l$ is the interferometer arm length. For simplicity, we consider the electron-tunneling regime of the beam splitters. The tunneling Hamiltonian is
\begin{equation}
H_T=\sum_{\sigma=\uparrow,\downarrow} \left(\mathcal{O}_{1\sigma}+\mathcal{O}_{2\sigma}
\right)+\mathrm{h.c.}
\label{eq:HT}
\end{equation}
$\mathcal{O}_{1\sigma}=\gamma_{1\sigma} e^{i\varphi_{\mathrm{AB}}^{\sigma}} e^{ieVt/\hbar} \psi_{\mathrm{P}\sigma}^{\dagger}(0,t) \psi_{\mathrm{N}\sigma}(0,t)$ and $\mathcal{O}_{2\sigma} = \gamma_{2\sigma}
e^{ieV(t-l/v_{\mathrm{N}c})/\hbar}
\psi_{\mathrm{P}\sigma}^{\dagger}(l,t)
\psi_{\mathrm{N}\sigma}(l,t)$ describe tunneling of spin-$\sigma$ electrons at the upper and lower
beam splitters, respectively. $\varphi_{\mathrm{AB}}^{\sigma}$ ($\sigma=\uparrow,\downarrow$)
is the AB phase associated with the loop formed by the spin-$\sigma$ channel in the p region and the
fractional channel in the n region (Fig.~\ref{fig4}b).
$\gamma_{i\sigma}$ are the tunneling strengths.
The applied voltage $V$ in the n region enters through the
Peierls phase factors $e^{ieVt/\hbar}$ and $e^{ieV(t-l/v_{\mathrm{N}c})/\hbar}$ (Supplementary Note~9).

The total Hamiltonian is $H=H_0+H_T$.
The current operator across the p-n junction is $I=(ie/\hbar) \sum_{\sigma=\uparrow,\downarrow}
\left(\mathcal{O}_{1\sigma} + \mathcal{O}_{2\sigma} - \mathrm{h.c.} \right)$.
We evaluate the nonequilibrium current perturbatively in
$H_T$. The current is $\langle I(t=0)\rangle
=\sum_{n,m\in\mathbb{Z}}I_{nm}$, where $I_{nm}$ denotes the contribution to the
interference frequency $f_{nm}$ proportional to the AB phase factor $e^{in\varphi_\AB^\ua} e^{im\varphi_\AB^\da}$. Computation of higher-order perturbation terms is required for evaluating $I_{nm}$ with larger $n$ and $m$.

\subsection*{Temperature dependence of interference}

We theoretically obtain the $f_{10}$, $f_{01}$, and $f_{11}$ components of the zero-bias conductance at $(\nu_{\mathrm{P}},\nu_{\mathrm{N}})=(-2,2/3)$ in the SU(2)-symmetric spin-unpolarized regime. Comparing their thermal decay with the experimental data in Fig.~\ref{fig3}e, we extract $\zeta$ in Eq.~\eqref{thermalvisibilities}. We assume $v_{\mathrm{N}c}>v_{\mathrm{P}\uparrow},v_{\mathrm{P}\downarrow}$, which is likely the case in the experiment, as repulsive interactions enhance the charge-mode velocity on the $\nu_{\mathrm{N}}=2/3$ edge~\cite{Hu2009}.

The $f_{10}$ component arises from the interference between the two subprocesses in Fig.~\ref{fig4}c. Its amplitude is determined by the overlaps between the corresponding excitations in the two interfering subprocesses: electrons on the integer channel, and chargons and spinons on the fractional channel. As these excitations propagate at different velocities, the three overlaps cannot all be maximal simultaneously. The largest contribution occurs when the chargon overlap is maximal, because the chargon has the largest scaling dimension. In this case, the corresponding electrons and spinons in the two subprocesses have relative time delays of $\Delta t_{\uparrow}=l/v_{\mathrm{P}\uparrow}-l/v_{\mathrm{N}c}$ and
$\Delta t_s=l/v_{\mathrm{N}s}+l/v_{\mathrm{N}c}$, respectively.
In the high-temperature regime
$k_{\mathrm B}T\gg\hbar/\Delta t_{\uparrow}, \,\hbar/\Delta t_s$, we obtain (Supplementary Note~5)
\begin{equation}
\begin{aligned}
\frac{\partial I_{10}}{\partial V}
\approx{}&
\frac{
8\pi^2 e^2
\gamma_{1\uparrow}
\gamma_{2\uparrow}^{*}
}{
\hbar^4
v_{\mathrm{P}\uparrow}
v_{\mathrm{N}c}^{3/2}
v_{\mathrm{N}s}^{1/2}
}
\,e^{i\varphi_{\mathrm{AB}}^{\uparrow}}
\,k_{\mathrm{B}}T
\\
&\times
\exp\!\left[
-\frac{\pi k_{\mathrm{B}}T}{\hbar}
\left(
\Delta t_{\uparrow}
+\frac{1}{2}\Delta t_s
\right)
\right].
\end{aligned}
\label{eq:I10thermal}
\end{equation}
Here, the coefficients $1$ and $1/2$ in the exponent are
twice the scaling dimensions of the spin-up electron and spinon operators, respectively. The prefactor $k_{\mathrm{B}}T$ is consistent with the Luttinger-liquid power law. At high temperatures, the exponential decay dominates.

The $f_{01}$ component arises from the same mechanism as $f_{10}$, with spin-down electrons and down-spinons replacing spin-up electrons and up-spinons. Its thermal decay is governed by $\Delta t_{\downarrow}=l/v_{\mathrm{P}\downarrow}-l/v_{\mathrm{N}c}$ and $\Delta t_s$. In the high-temperature regime $k_{\mathrm B}T\gg\hbar/\Delta t_{\downarrow},\,\hbar/\Delta t_s$,
\begin{equation}
\begin{aligned}
\frac{\partial I_{01}}{\partial V}
\approx{}&
\frac{
8\pi^2e^2
\gamma_{1\downarrow}
\gamma_{2\downarrow}^{*}
}{
\hbar^4
v_{\mathrm{P}\downarrow}
v_{\mathrm{N}c}^{3/2}
v_{\mathrm{N}s}^{1/2}
}
\,e^{i\varphi_{\mathrm{AB}}^{\downarrow}}
\,k_{\mathrm B}T
\\
&\times
\exp\!\left[
-\frac{\pi k_{\mathrm B}T}{\hbar}
\left(
\Delta t_{\downarrow}
+\frac{1}{2}\Delta t_s
\right)
\right].
\end{aligned}
\label{eq:I01thermal}
\end{equation}

The $f_{11}$ component arises at higher order in the perturbation theory (Supplementary Note~6). The dominant contribution to the interference arises when a spin-up electron and a spin-down electron tunnel nearly simultaneously, within the thermal time scale $O(\hbar/k_{\mathrm B}T)$, from the integer channels to the fractional channel at the same beam splitter (Fig.~\ref{fig4}d). 
The chargons maximally overlap between the two interfering subprocesses, while the up-spinon and the down-spinon fuse into the vacuum channel in each subprocess, following Eq.~\eqref{spinonfusion}. The spinon fusion is described by the operator product expansion
\begin{equation} \label{spinonOPE}
	e^{i\sqrt{\frac12}\phi_{\mathrm{N}s}(x,t)}e^{-i\sqrt{\frac12}\phi_{\mathrm{N}s}(x,t')}\propto [a+i(t-t')]^{-1/2}\,\mathbbm 1+\cdots,
\end{equation}
whose leading term is the identity operator $\mathbbm 1$.
The temporal delays between the two interfering subprocesses are associated with the spin-up and spin-down electrons on the integer channels in the p region, given by $\Delta t_{\uparrow}$ and $\Delta t_{\downarrow}$, respectively.
In the high-temperature regime,
$k_{\mathrm B}T\gg\hbar/\Delta t_{\downarrow},
\,\hbar/\Delta t_{\uparrow}, \, \hbar/(l/v_{\mathrm Nc})$, we obtain
\begin{equation}
\begin{aligned}
\frac{\partial I_{11}}{\partial V}
\approx{}&
-\frac{i e^2
\gamma_{1\uparrow}
\gamma_{1\downarrow}
\gamma_{2\uparrow}^{*}
\gamma_{2\downarrow}^{*}}
{\hbar^7
v_{\mathrm{P}\uparrow}
v_{\mathrm{P}\downarrow}
v_{\mathrm{N}c}^3
v_{\mathrm{N}s}}
\,e^{i(\varphi_{\mathrm{AB}}^{\uparrow}
+\varphi_{\mathrm{AB}}^{\downarrow})}
\\
&\times
(2C)\,
(k_{\mathrm B}T)^2
\exp\!\left[
-\frac{\pi k_{\mathrm B}T}{\hbar}
\left(
\Delta t_{\uparrow}
+
\Delta t_{\downarrow}
\right)
\right],
\end{aligned}
\label{eq:I11thermal}
\end{equation}
where $C\approx 7938\,i$ is a dimensionless constant. The exponential decay factor is free of spinon-induced decoherence due to the fusion of the up-spinon and down-spinon into the vacuum channel. From Eqs.~\eqref{eq:I10thermal}, \eqref{eq:I01thermal}, \eqref{eq:I11thermal} and $\mathcal V_{nm} \propto |\partial I_{nm}/\partial V|$, we obtain Eq.~\eqref{thermalvisibilities} with $\zeta = \pi k_{\mathrm B} \Delta t_s/(2\hbar)$.

Comparing the results in Eqs.~\eqref{eq:I10thermal}, \eqref{eq:I01thermal}, \eqref{eq:I11thermal} with the experimental data in Fig.~\ref{fig3}e, we estimate $\Delta t_{\uparrow}=\SI{7.31}{ps}$, $\Delta t_{\downarrow}=\SI{1.67}{ps}$, and $\Delta t_s=\SI{2.18}{ps}$ (Supplementary Note~7). 
The estimated value of $\Delta t_s=\SI{2.18}{ps}$ corresponds to $\zeta=\SI{0.45}{K^{-1}}$.

We note that the temperature dependence of the interferences can be compared with their bias dependence in Extended Data Fig.~\ref{lobes}.

\textbf{\begin{center}Author Contributions\end{center}}

P.R. and H.-S.S. co-supervised the project. R.A., M.K., and Q.B. fabricated the devices and performed the measurements. K.K. and H.-S.S. developed the theoretical framework. R.A., M.K., Q.B., H.C., L.P., K.K., H.-S.S., and P.R. participated in the analysis and interpretation of the data. K.W. and T.T. provided the hBN crystals. R.A., K.K., H.-S.S., and P.R. wrote the manuscript with input from all authors.

\textbf{\begin{center}Acknowledgements\end{center}}

This work was supported by the European Research Council (ERC) Starting Grant COHEGRAPH (Grant No.~679531, P.R.), the Agence Nationale de la Recherche (ANR) through the EQUBITFLY project (P.R.), the Horizon Europe EIC Pathfinder Open project FLATS (Grant No.~101130384, P.R.), the Horizon Europe EIC Pathfinder Open project ELEQUANT (P.R.), and Korea NRF (Grant No. RS-2026-25469730, H.-S.S.). 

\textbf{\begin{center}Competing interests\end{center}}
The authors declare no competing interests.

\textbf{\begin{center}Data availability\end{center}}
The data supporting the findings of this study are available from the corresponding author upon reasonable request.


\bibliography{Inter}

\newpage
\clearpage
\onecolumngrid

\section*{Extended Data}
\renewcommand{\thefigure}{ED\arabic{figure}}
\setcounter{figure}{0}
\setcounter{equation}{0}

\begin{figure*}[th]
  \centering
  \includegraphics[width=1\textwidth]{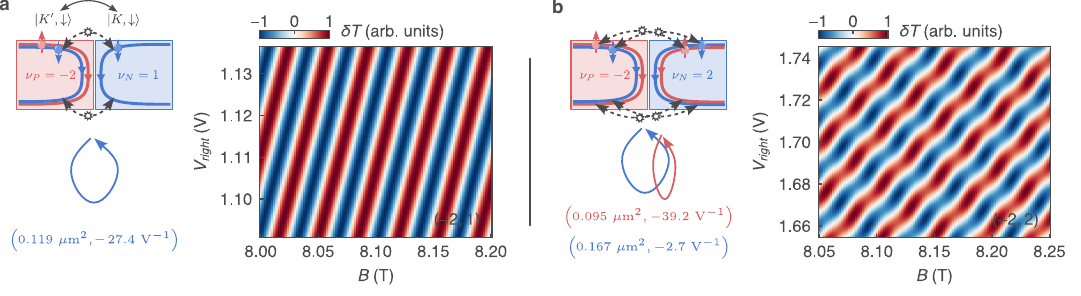}
  \caption{\textbf{Simulations of the interferometric response in Fig.~1e,f.}
  \textbf{a,b,} Simulated AB oscillations of the differential transmission $\delta \mathcal{T}$ as a function of magnetic field $B$ and right-gate voltage $V_{\mathrm{right}}$, corresponding to the experimental data shown in Fig.~1e,f. \textbf{a,} Simulation for $(\nu_{\mathrm{P}},\nu_{\mathrm{N}})=(-2,1)$, where a single spin-down Mach--Zehnder interferometer (MZI) contributes to the interference pattern. \textbf{b,} Simulation for $(\nu_{\mathrm{P}},\nu_{\mathrm{N}})=(-2,2)$, where spin-down and spin-up MZIs coexist and jointly determine the interference pattern.}

  \label{simu}
\end{figure*}

\begin{figure*}[th]
  \centering
  \includegraphics[width=1\textwidth]{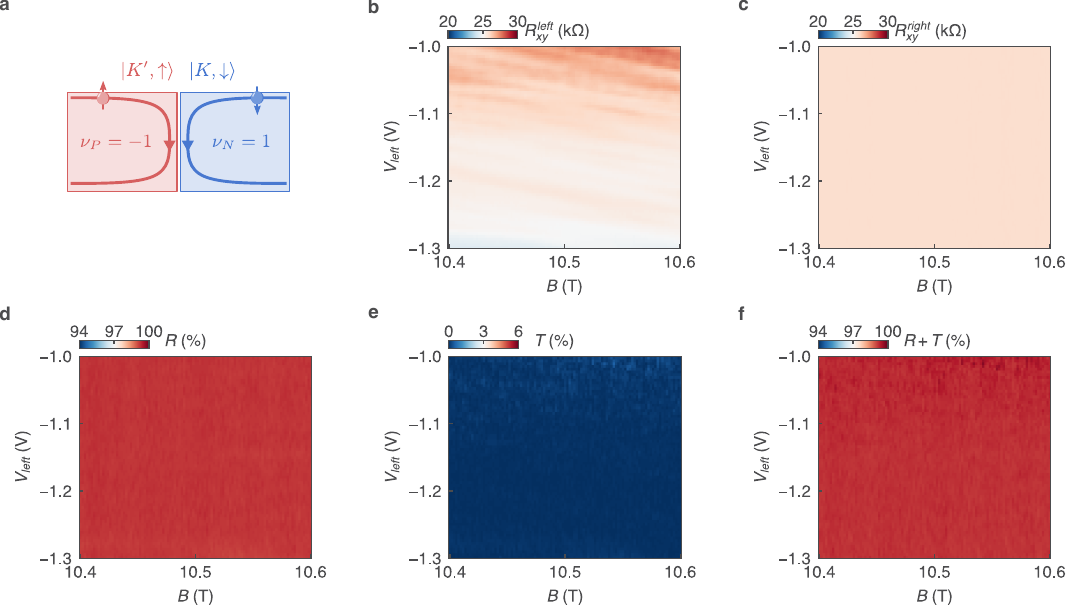}
  \caption{\textbf{Absence of interchannel mixing at $\mathbf{(\nu_P,\nu_N)=(-1,1)}$.}
  \textbf{a,} Edge-channel configuration for $(\nu_{\mathrm{P}},\nu_{\mathrm{N}})=(-1,1)$.
  \textbf{b,c,} Two-point Hall resistances of the left $R_{xy}^{\mathrm{left}}$ (\textbf{b}) and right $R_{xy}^{\mathrm{right}}$ (\textbf{c}) graphene halves as functions of left-gate voltage $V_{\mathrm{left}}$ and magnetic field $B$.
  \textbf{d--f,} Reflection $\mathcal{R}$ (\textbf{d}), transmission $\mathcal{T}$ (\textbf{e}), and their sum $\mathcal{R}+\mathcal{T}$ (\textbf{f}), as functions of $V_{\mathrm{left}}$ and $B$.
  Nearly all of the injected current is reflected at the junction, demonstrating the absence of mixing between the two copropagating edge channels with opposite spin and valley polarizations.
}
  \label{figED-1_1}
\end{figure*}

\begin{figure*}[th]
  \centering
  \includegraphics[width=1\textwidth]{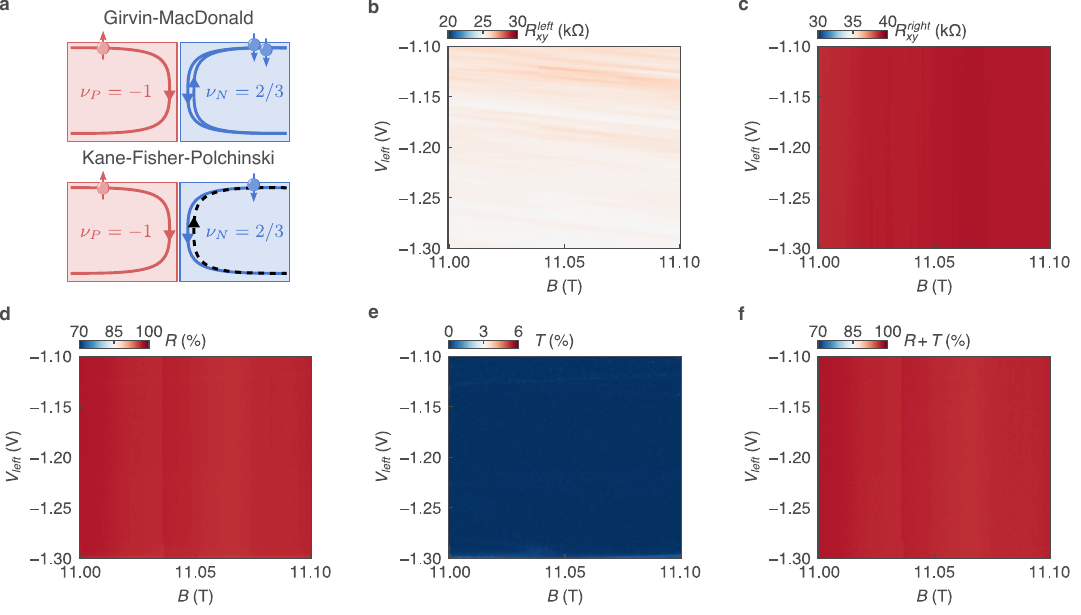}
  \caption{\textbf{Absence of interchannel mixing at $\mathbf{(\nu_{\mathrm{P}},\nu_{\mathrm{N}})=(-1,2/3)}$ in the spin-polarized $\mathbf{\nu=2/3}$ regime.}
  \textbf{a,} Edge-channel configurations for $(\nu_{\mathrm{P}},\nu_{\mathrm{N}})=(-1,2/3)$, shown within both the Girvin--MacDonald and Kane--Fisher--Polchinski (KFP) descriptions of the $\nu=2/3$ edge.
  \textbf{b,c,} Two-point Hall resistances of the left $R_{xy}^{\mathrm{left}}$ (\textbf{b}) and right $R_{xy}^{\mathrm{right}}$ (\textbf{c}) graphene halves as functions of left gate voltage $V_{\mathrm{left}}$ and magnetic field $B$.
  \textbf{d--f,} Reflection $\mathcal{R}$ (\textbf{d}), transmission $\mathcal{T}$ (\textbf{e}), and their sum $\mathcal{R}+\mathcal{T}$ (\textbf{f}), as functions of $V_{\mathrm{left}}$ and $B$.
  Nearly all of the injected current is reflected at the junction, demonstrating the absence of mixing between the copropagating edge channels across the junction when the $\nu=2/3$ state is spin-polarized.
  }
  \label{figED-1_23}
\end{figure*}

\begin{figure*}[th]
    \centering
    \includegraphics[width=1\textwidth]{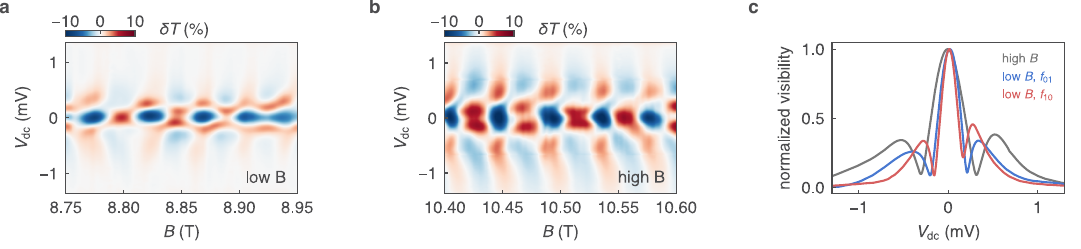}
    \caption{\textbf{Bias-voltage dependence of the interference at $\mathbf{(\nu_{\mathrm{P}},\nu_{\mathrm{N}})=(-2,2/3)}$ in the spin-unpolarized and spin-polarized $\mathbf{\nu=2/3}$ regimes.}
    \textbf{a,b,} Bias-voltage dependence of the Aharonov--Bohm oscillations at $(\nu_{\mathrm{P}},\nu_{\mathrm{N}})=(-2,2/3)$ in the low- (\textbf{a}) and high- (\textbf{b}) magnetic-field regimes.
    \textbf{c,} Interference visibilities normalized to their respective zero-bias values, for the $f_{10}$ and $f_{01}$ components in the low-field regime and for the single frequency component observed in the high-field regime.
    }
\label{lobes}
\end{figure*}

\clearpage

\begin{center}
{\Large\bfseries Supplementary Information}\\[4mm]

{\Large\bfseries
Interferometry reveals spin-singlet fractional quantum Hall edges
}\\[8mm]
\setcounter{page}{1}
\thispagestyle{empty}

R. Ayache$^{1,\dagger}$,
K. Kim$^{2,\dagger}$,
M. Kuiri$^{1,3,\dagger}$,
Q. Benichou$^{1}$,
H. Chakraborti$^{1}$,
L. Pugliese$^{1}$,
K. Watanabe$^{4}$,
T. Taniguchi$^{4}$,
H.-S. Sim$^{2,*}$,
P. Roulleau$^{1,*}$

\vspace{4mm}

{\small
$^{1}$SPEC, CEA, CNRS, Université Paris-Saclay,
CEA Saclay, 91191 Gif-sur-Yvette Cedex, France\\[1mm]

$^{2}$Department of Physics,
Korea Advanced Institute of Science and Technology,
Daejeon 34141, Republic of Korea\\[1mm]

$^{3}$Department of Physics, Birla Institute of Technology and Science, Pilani, Hyderabad Campus, Jawahar Nagar, Kapra Mandal, Medchal District, Telangana 500078, India\\[1mm]

$^{4}$National Institute for Materials Science,
1-1 Namiki, Tsukuba 305-0044, Japan\\[2mm]

$^{\dagger}$These authors contributed equally.\\
$^{*}$Correspondence:
\texttt{hs\_sim@kaist.ac.kr};
\texttt{preden.roulleau@cea.fr}
}
\end{center}

\vspace{8mm}
\supplementarytableofcontents
\newpage

\setcounter{equation}{0}
\renewcommand{\theequation}{S\arabic{equation}}
\renewcommand{\theHequation}{S\arabic{equation}}
\setcounter{figure}{0}
\renewcommand{\thefigure}{S\arabic{figure}}
\renewcommand{\theHfigure}{S\arabic{figure}}

\supplementarynote{1}{Device fabrication and geometry}
\label{SN1}

The devices were fabricated from high-quality hBN/graphene/hBN/graphite van der Waals heterostructures assembled using a dry-transfer technique. Monolayer graphene and few-layer graphite flakes were mechanically exfoliated from natural graphite (NGS GmbH) onto Si/SiO$_2$ substrates with a \SI{285}{nm} oxide layer. Hexagonal boron nitride (hBN) crystals supplied by NIMS were exfoliated onto separate substrates.

Suitable flakes were first identified by optical microscopy and further inspected using dark-field and differential-interference-contrast imaging to exclude visible cracks and thickness inhomogeneities. Non-contact atomic force microscopy (AFM) was then employed to verify surface cleanliness and thickness uniformity before assembly.

\begin{figure}[htbp]
\centering
\includegraphics[width=\textwidth]{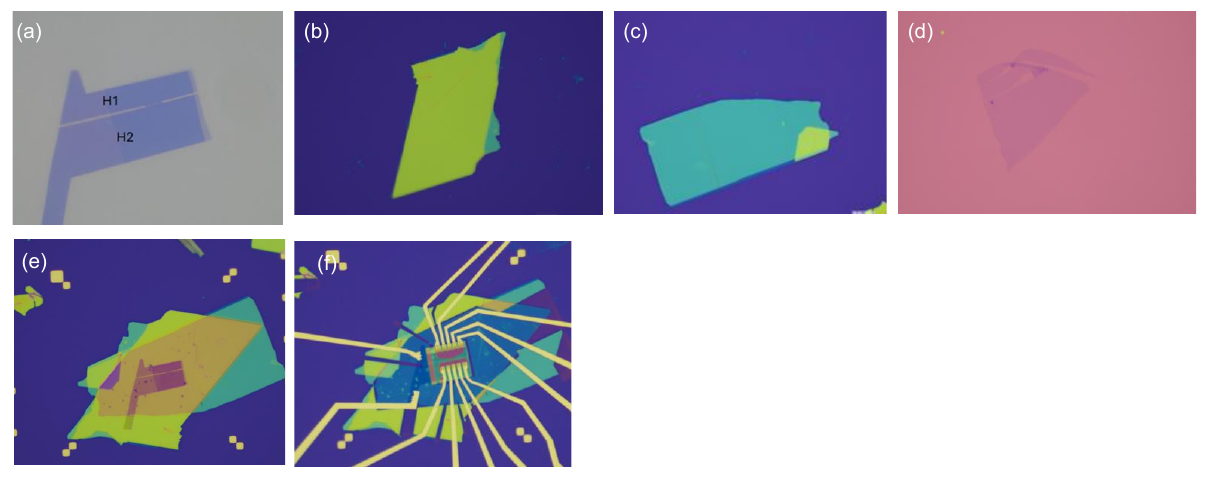}
\caption{\textbf{Optical images of the fabrication steps.} (a) Bottom graphite gate split into two halves, with a separation of approximately \SI{70}{nm}. (b)--(d) Bottom hBN, top hBN, and monolayer graphene, respectively. (e) Complete stack, consisting of graphite split gates/hBN/graphene/hBN. (f) Final device after patterning and contact deposition.}
\label{sfig:figS1}
\end{figure}

A selected graphite flake was transferred onto a pre-patterned Si/SiO$_2$ substrate using a polypropylene carbonate (PC) / polydimethylsiloxane (PDMS) stamp. Electron-beam lithography followed by oxygen plasma etching defined two independent graphite bottom gates, referred to as the left and right gates, separated by a $\sim \SI{70}{nm}$-wide gap [Fig.~\ref{sfig:figS1}(a)] and a width of $\sim \SI{0.7}{\micro\meter}$. After lithography, the substrate was cleaned in hot acetone and annealed in vacuum at \SI{350}{\celsius} to remove fabrication residues before a second AFM inspection.

The active heterostructure, consisting of monolayer graphene encapsulated between two hBN flakes (approximately \SI{37}{nm} on the top and \SI{60}{nm} on the bottom), was subsequently assembled and accurately aligned onto the patterned graphite gates [Figs.~\ref{sfig:figS1}(b)--(e)]. A second cleaning step and vacuum annealing at \SI{350}{\celsius} were performed to improve the interface quality.

The graphene stack was patterned by electron-beam lithography and CHF$_3$/O$_2$ reactive-ion etching. One-dimensional edge contacts were fabricated by exposing the graphene edges with a second etching step, followed by angled electron-beam evaporation of Cr/Au (10/\SI{70}{nm}). The contacts were designed with a finger geometry to minimize the contact resistance. Finally, the graphene channel was defined by an additional lithography and etching step. The optical image of the final device is shown in Fig.~\ref{sfig:figS1}(f). 


\supplementarynote{2}{Device geometry}
\label{SN2}

\begin{figure}[htbp]
\centering
\includegraphics[width=0.6\textwidth]{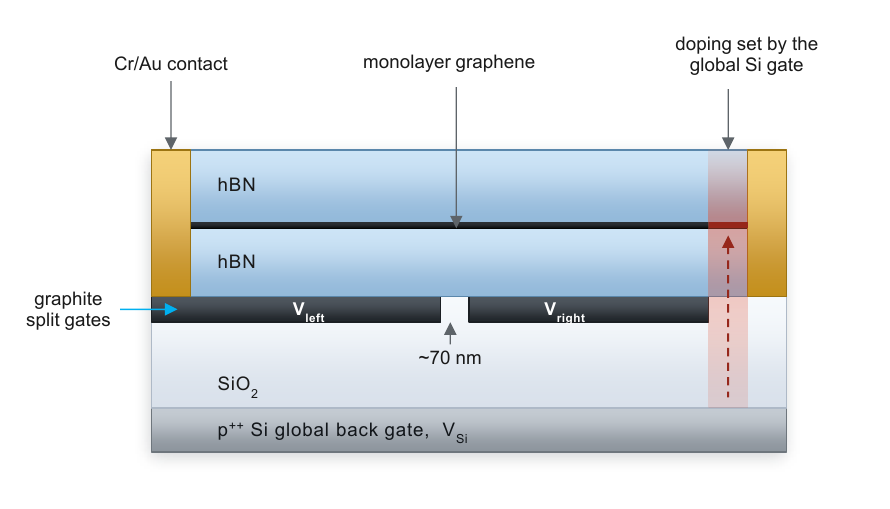}
\caption{Schematic of the device structure.}
\label{sfig:figS2}
\end{figure}

The schematic of the device geometry is shown in Fig.~\ref{sfig:figS2}, which consists of two independent graphite gates labeled as $V_{\mathrm{left}}$ and $V_{\mathrm{right}}$, separated by a lateral distance of $\sim \SI{70}{nm}$. On the left side, the electrical contacts are entirely located above the local graphite gate, whereas on the right side they partially extend onto the SiO$_2$/Si substrate. This geometry allows the right contact region to be additionally electrostatically tuned by the global silicon back gate, thereby reducing the contact resistance. Voltages of opposite polarities are applied to $V_{\mathrm{left}}$ and $V_{\mathrm{right}}$ gates independently to create a p-n junction. Additional gate voltage, $V_{\mathrm{Si}}$ is used on the right side to tune contact resistance of graphene on the right region.

\supplementarynote{3}{Transport measurement}
\label{SN3}

Electrical transport measurements were carried out in an Oxford Instruments Proteox dilution refrigerator operating at a base temperature of approximately \SI{10}{mK} with a perpendicular magnetic field up to \SI{14}{T}.
The device response was measured using standard low-frequency lock-in techniques. An AC excitation current between 1 and \SI{5}{nA} was supplied by a Stanford Research Systems SR830 lock-in amplifier. Depending on the measurement configuration, modulation frequencies around \SI{7}{Hz} were employed to optimize the signal-to-noise ratio.
The resulting voltage signals were amplified using a low-noise, high-input-impedance preamplifier with an input impedance of approximately \SI{100}{\mega\ohm}. To suppress electrical noise and prevent unwanted electron heating, all measurement lines were equipped with low-pass RC filters and thermally anchored at several temperature stages throughout the dilution refrigerator.

\supplementarynote{4}{Model for the spin-polarized regime of $(\nu_{\mathrm{P}},\nu_{\mathrm{N}})=(-2,2/3)$}
\label{SN4}

We present a model for the MZI formed along the $(\nu_{\mathrm{P}},\nu_{\mathrm{N}})=(-2,2/3)$ p-n interface in the spin-polarized regime in Fig.~4a of the main text. The MZI arms are composed of the interface channels. In the $\nu_{\mathrm{P}}=-2$ region, there appear spin-up and spin-down integer channels. In the $\nu_{\mathrm{N}}=2/3$ region, there are an upstream charge channel between the $\nu_{\mathrm{N}}=2/3$ region and the $\nu=1$ interface region, and a downstream charge channel between the $\nu=1$ and $\nu=0$ interface regions~\cite{Macdonald2over3}; the downstream channel has the same character as an integer QH channel. The two counterpropagating charge channels in the n region carry spin down, and they are spatially separated by the $\nu=1$ interface region, where there is no disorder causing the interchannel electron tunneling (see the main text). The Hamiltonian $H_0^{\textrm{pl}}$ for the MZI arms is written as 
$$H_0^{\textrm{pl}} = \sum_{\sigma\in\{\ua,\da\}} \frac{-i \hbar v_{\mathrm{P}\sigma}}{2\pi}\int dx\: \psi_{\mathrm{P}\sigma}^\dagger \partial_x \psi_{\mathrm{P}\sigma} + \sum_{\alpha\in\{d,u\}} \frac{\hbar v_{\mathrm{N}\alpha}}{4\pi}\int dx\:(\partial_x\phi_{\mathrm{N}\alpha})^2.$$
Here $\psi_{\mathrm{P}\sigma}$ describes spin-$\sigma$ electrons in the $\nu_{\mathrm{P}}=-2$ region ($\sigma=\uparrow,\downarrow$), while $\phi_{\mathrm{N}d}$ and $\phi_{\mathrm{N}u}$ describe the downstream and upstream charge channels in the $\nu_{\mathrm{N}}=2/3$ region, respectively. They satisfy $[\phi_{\mathrm{N}d}(x_1), \phi_{\mathrm{N}d}(x_2)] = i\pi \sgn(x_1-x_2)$, $[\phi_{\mathrm{N}u}(x_1), \phi_{\mathrm{N}u}(x_2)] = -i\pi \sgn(x_1-x_2)$, and $[\phi_{\mathrm{N}d}(x_1), \phi_{\mathrm{N}u}(x_2)]=0$.

We next describe the upper and lower beam splitters, which are located at $x=0$ and $l$, respectively. Since the n region is spin polarized in the spin-down direction, only spin-down electrons can move between the p and n regions at the beam splitters. We assume that only the downstream charge channel in the n region couples to the spin-down integer channel in the p region via the beam splitters, and that interchannel Coulomb interactions are negligible. Then the Hamiltonian for the beam splitters is written as
\begin{equation} \label{HTpl}
	H_T^{\textrm{pl}}(t) = \gamma_1 e^{i\varphi_\AB} e^{ieVt/\hbar} \psi_{\mathrm{P}\da}^\dagger(0,t) \psi_{\mathrm{N}d}(0,t) + \gamma_2 e^{ieV(t-l/v_{\mathrm{N}d})/\hbar} \psi_{\mathrm{P}\da}^\dagger(l,t) \psi_{\mathrm{N}d}(l,t) + \text{h.c.}
\end{equation}
Here, $\gamma_1$ and $\gamma_2$ are the tunneling strengths at the upper and lower beam splitters, respectively. $\varphi_{\AB}$ is the AB phase of spin-down electrons in the MZI, and the operator $\psi_{\mathrm{N}d}\propto e^{-i\phi_{\mathrm{N}d}}$ describes an electron in the downstream charge channel of the n region. The applied voltage $V$ in the n region is incorporated through the Peierls phase factors $e^{ieVt/\hbar}$ and $e^{ieV(t-l/v_{\mathrm{N}d})/\hbar}$. The total Hamiltonian is $H^{\textrm{pl}}=H_0^{\textrm{pl}}+H_T^{\textrm{pl}}(t)$.

This Hamiltonian describes an MZI of noninteracting spin-down electrons. The interference visibility of the differential conductance through the MZI is obtained as
\begin{equation}
	\mathcal{V} = \frac{2\sqrt{\mathcal{T}_1\mathcal{T}_2\mathcal{R}_1\mathcal{R}_2}}{\mathcal{T}_1\mathcal{R}_2+\mathcal{T}_2\mathcal{R}_1} \frac{\pi k_BT \Delta t/\hbar}{\sinh(\pi k_BT\Delta t/\hbar)},
\end{equation}
where $\Delta t=\frac{l}{v_{\mathrm{N}d}}-\frac{l}{v_{\mathrm{P}\downarrow}}$, $v_{\mathrm{P}\downarrow}$ is the velocity of a spin-down electron in the spin-down integer channel of the p region, and $v_{\mathrm{N}d}$ is the velocity of a downstream spin-down channel in the n region. The transmission coefficients are written in terms of the tunneling amplitudes $\gamma_1$ and $\gamma_2$ as $\mathcal{T}_i =\frac{4\pi^2|\gamma_i|^2/\hbar^2 v_{\mathrm{P}\da}v_{\mathrm{N}d}}{(1+\pi^2|\gamma_i|^2/\hbar^2v_{\mathrm{P}\da}v_{\mathrm{N}d})^2}$, and $\mathcal{R}_i=1-\mathcal{T}_i$. In the high-temperature regime $k_BT\gg\frac{\hbar}{|\Delta t|}$, this expression is reduced to $\mathcal{V}\propto Te^{-\pi k_BT|\Delta t|/\hbar}$.

\supplementarynote{5}{Conductance by (1,0) and (0,1) interferences}
\label{SN5}

We derive Eq.~\eqref{eq:I10thermal} of the main text, the conductance $\frac{\partial I_{10}}{\partial V}$ by the $(1,0)$ interference at high temperatures $k_BT \gg \frac{\hbar}{\Delta t_\ua}, \frac{\hbar}{\Delta t_s}$. Recall that $I_{nm}$ denotes the current contribution proportional to $e^{i(n\varphi_\AB^\ua + m\varphi_\AB^\da)}$, corresponding to $n$ and $m$ windings around the spin-up and spin-down AB loops, respectively. Hereafter we set $\hbar \equiv 1$ and $k_B \equiv 1$ sometimes.

We start from the current operator $I(t)=ie\sum_{\sigma\in\{\ua,\da\}}\left[\O_{1\sigma}(t)+\O_{2\sigma}(t)- \textrm{h.c.}\right]$ (Methods of the main text).
The contribution of the $(1,0)$ interference process to the current flowing from the n region to the p region has the form
$$I_{10} = -e \int_{-\infty}^{\infty} dt\, \left(\langle \O^\dagger_{2\ua}(t)\,\O_{1\ua}(0)\rangle-\langle \O_{1\ua}(t)\,\O^\dagger_{2\ua}(0)\rangle \right),$$
capturing the interference between the two subprocesses in which tunneling occurs at the upper and lower beam splitters, respectively (Fig.~4c of the main text). The first correlator becomes
\[\begin{aligned}
	\langle \O_{2\ua}^\dagger(t) \O_{1\ua}(0) \rangle
	&= \frac{\gamma_{1\ua} \gamma_{2\ua}^*}{a^2} e^{i\varphi_\AB^\ua} e^{-ieV\left(t-\frac{l}{v_{\mathrm{N}c}}\right)}   \left\langle \psi_{\mathrm{P}\ua}(l,t) \psi_{\mathrm{P}\ua}^\dagger(0,0) \right\rangle
	\left\langle e^{i\sqrt{\frac32}\phi_{\mathrm{N}c}(l,t)} e^{-i\sqrt{\frac32}\phi_{\mathrm{N}c}(0,0)} \right\rangle
	\left\langle e^{i\sqrt{\frac12}\phi_{\mathrm{N}s}(l,t)} e^{-i\sqrt{\frac12}\phi_{\mathrm{N}s}(0,0)} \right\rangle \\
	&= \frac{\gamma_{1\ua} \gamma_{2\ua}^*}{v_{\mathrm{P}\ua} v_{\mathrm{N}c}^{3/2} v_{\mathrm{N}s}^{1/2}} e^{i\varphi_\AB^\ua} e^{-ieV\left(t-\frac{l}{v_{\mathrm{N}c}}\right)} K(t).
\end{aligned}\]
where $K(t) = G_{-1}\left(t- \frac{l}{v_{\mathrm{P}\ua}}\right) G_{-3/2}\left(t-\frac{l}{v_{\mathrm{N}c}}\right) G_{-1/2}\left(t+\frac{l}{v_{\mathrm{N}s}}\right)$ and $G_{-\alpha}(t) = \left( \frac{i}{\pi T} \sinh\left[ \pi T (t - ia)\right] \right)^{-\alpha}$ with a short-distance cutoff $a=0^+$~\cite{bosonization}. 
The three factors in $K(t)$ arise from the overlaps between the corresponding excitations (see Fig.~4c) of the $\nu_{\mathrm{P}}=-2$ spin-up channel, the $\nu_{\mathrm{N}}=2/3$ charge mode, and the $\nu_{\mathrm{N}}=2/3$ spin mode, each generated by the tunneling in the two subprocesses.
The associated exponents, $\alpha=1$, $3/2$, and $1/2$, are twice the scaling dimensions of these respective excitations.
Evaluating the second correlator in the same way, the differential conductance at $V=0$ is found as
\begin{equation} \label{dIdV10}
	\frac{\partial I_{10}}{\partial V}
	= \frac{ie^2\,\gamma_{1\ua}\gamma_{2\ua}^*}{v_{\mathrm{P}\ua} v_{\mathrm{N}c}^{3/2} v_{\mathrm{N}s}^{1/2}}e^{i\varphi_\AB^\ua}
	\left[
	\int_{-\infty}^{\infty} dt\,\left(t-\frac{l}{v_{\mathrm{N}c}}\right)K(t)
	-\textrm{c.c.}
	\right].
\end{equation}
In the high-temperature regime, each factor of $K(t)$ exhibits exponential decay $G_{-\alpha}(t)\sim e^{-\alpha\pi T|t|}$ for $|t| \gtrsim 1/T$. The integral is therefore dominated by the maximum of the exponent in $K(t)\sim e^{\pi T S(t)}$, where
$$S(t)= -\left| t-\frac{l}{v_{\mathrm{P}\ua}} \right|
-\frac32\left| t-\frac{l}{v_{\mathrm{N}c}} \right|
-\frac12\left| t+\frac{l}{v_{\mathrm{N}s}} \right|$$
measures the total temporal mismatch between the corresponding excitations of the two interfering subprocesses, weighted by their exponents. For $v_{\mathrm{N}c}>v_{\mathrm{P}\ua}$, $S(t)$ has a unique maximum at $t=l/v_{\mathrm{N}c}$. At this point, the excitations of the $\nu_{\mathrm{N}}=2/3$ charge mode, which has the largest exponent, have the perfect overlap between the two subprocesses, while those of the $\nu_{\mathrm{P}}=-2$ spin-up channel and the $\nu_{\mathrm{N}}=2/3$ spin mode exhibit the temporal mismatches $\Delta t_\ua = \frac{l}{v_{\mathrm{P}\ua}} - \frac{l}{v_{\mathrm{N}c}}$ and $\Delta t_s = \frac{l}{v_{\mathrm{N}s}} + \frac{l}{v_{\mathrm{N}c}}$, respectively, between the subprocesses. The maximum value of $S(t)$ is $S_* = -\left(\Delta t_\ua + \frac12 \Delta t_s\right)$, giving the leading exponential suppression $\sim e^{\pi T S_*}$ of the integral in Eq.~\eqref{dIdV10}.

To extract the prefactor, we consider fluctuations around $t=l/v_{\mathrm{N}c}$ by introducing $u = \pi T (t-\frac{l}{v_{\mathrm{N}c}})$. Then Eq.~\eqref{dIdV10} becomes
\begin{equation} \label{10result}
	\int_{-\infty}^{\infty} dt\,\left(t-\frac{l}{v_{\mathrm{N}c}}\right)K(t)
	\approx
	C\, T \, e^{-\pi T\left(\Delta t_\ua + \frac12 \Delta t_s\right)},
\end{equation}
where the dimensionless constant is
$$C = 2^{3/2}\pi e^{i\pi/4}\int_{-\infty}^{\infty} du \,
\frac{ue^{u/2}}{\left[i\sinh(u-ia)\right]^{3/2}} = 8\pi\log 2 - 4i\pi^2.$$
To obtain Eq.~\eqref{10result}, we used, e.g., $G_{-1}\left(t-\frac{l}{v_{\mathrm{P}\ua}}\right) \approx (2\pi T)i e^{-\pi T \Delta t_\ua} e^u$ in the expression of $K(t)$. In the evaluation of $C$, it is useful to shift $u\to u-i\pi/2$ in the complex plane and use $\sinh(X-i\pi/2) = -i\cosh X$. Combining these results, we obtain Eq.~\eqref{eq:I10thermal} of the main text,
\begin{equation}
	\frac{\partial I_{10}}{\partial V}
	\approx
	\frac{8\pi^2 e^2\gamma_{1\ua}\gamma_{2\ua}^*}{v_{\mathrm{P}\ua} v_{\mathrm{N}c}^{3/2} v_{\mathrm{N}s}^{1/2}}e^{i\varphi_\AB^\ua}\, T\, e^{-\pi T\left(\Delta t_\ua + \frac12 \Delta t_s\right)}.
\end{equation}

\supplementarynote{6}{Conductance by (1,1) interference}
\label{SN6}

We derive Eq.~\eqref{eq:I11thermal} of the main text, the conductance by the $(1,1)$ interference, to the leading nonvanishing order in the tunneling at high temperatures $k_BT \gg \frac{\hbar}{\Delta t_\ua}, \frac{\hbar}{\Delta t_\da}, \frac{\hbar}{l/v_{\mathrm{N}c}}$. Here $\Delta t_\sigma = \frac{l}{v_{\mathrm{P}\sigma}} - \frac{l}{v_{\mathrm{N}c}}$, $\sigma \in \{\ua, \da\}$ and $\Delta t_s = \frac{l}{v_{\mathrm{N}s}} + \frac{l}{v_{\mathrm{N}c}}$.

We start from the current operator $I(t)=ie\sum_{\sigma\in\{\ua,\da\}}\left[\O_{1\sigma}(t)+\O_{2\sigma}(t)- \textrm{h.c.}\right]$ (Methods of the main text). 
The contribution to the current by the $(1,1)$ interference is expressed as
$$I_{11}=ie\sum_{\sigma\in\{\ua,\da\}}\left[\langle \O_{1\sigma}(0)\rangle_{11}-\langle \O_{2\sigma}^\dagger(0)\rangle_{11}\right],$$
where $\langle\cdots\rangle_{11}$ means that only the contributions proportional to the phase factor $e^{i(\varphi_\AB^\ua+\varphi_\AB^\da)}$ (which represent the $(1,1)$ interference) are retained in the perturbative expansion. We note that $\langle \O_{1\sigma}^\dagger(0) \rangle$ and $\langle \O_{2\sigma}(0) \rangle$ instead give the complex-conjugate contribution, proportional to $e^{-i(\varphi_\AB^\ua+\varphi_\AB^\da)}$. 

We evaluate $\langle \O_{1\ua}(0)\rangle_{11}$ and $\langle \O_{2\ua}^\dagger(0)\rangle_{11}$ using the Keldysh perturbation theory; the spin-down terms $\langle \O_{1\da}(0)\rangle_{11}$ and $\langle \O_{2\da}^\dagger(0)\rangle_{11}$ are obtained directly from $\langle \O_{1\ua}(0)\rangle_{11}$ and $\langle \O_{2\ua}^\dagger(0)\rangle_{11}$, respectively, by exchanging the spin labels on the $\nu_{\mathrm{P}}=-2$ edge, i.e. $v_{\mathrm{P}\ua} \leftrightarrow v_{\mathrm{P}\da}$. The leading nonvanishing contribution appears at fourth order in the tunneling,
\begin{equation} \label{eq:O1u_keldysh}
	\langle \O_{1\ua}(0)\rangle_{11}=i\sum_{\vec\eta}\eta_1\eta_2\eta_3\int_{t_1,t_2,t_3<0}\left\langle T_K\,\O_{1\ua}(0^+)\O_{1\da}(t_1^{\eta_1})\O_{2\ua}^\dagger(t_2^{\eta_2})\O_{2\da}^\dagger(t_3^{\eta_3})\right\rangle
\end{equation}
where $T_K$ denotes ordering along the Keldysh contour and $\vec\eta=(\eta_1,\eta_2,\eta_3)$ with $\eta_j=\pm1$ indicates whether $t_j$ lies on the forward ($+$) or backward ($-$) branch of the contour. We use the shorthand $\int_{t_1,t_2,t_3<0}\equiv\int_{-\infty}^0 dt_1\int_{-\infty}^0 dt_2\int_{-\infty}^0 dt_3$. This term contributes to the $(1,1)$ interference as $\O_{1\ua}\O_{1\da} \O_{2\ua}^\dagger \O_{2\da}^\dagger \propto e^{i(\varphi_\AB^\ua+\varphi_\AB^\da)}$. Similarly, we obtain
\begin{equation}\label{eq:O2u_keldysh}
	\langle \O_{2\ua}^\dagger(0)\rangle_{11}=i\sum_{\vec\eta}\eta_1\eta_2\eta_3\int_{t_1,t_2,t_3<0}\left\langle T_K\,\O_{2\ua}^\dagger(0^+)\O_{2\da}^\dagger(t_1^{\eta_1})\O_{1\ua}(t_2^{\eta_2})\O_{1\da}(t_3^{\eta_3})\right\rangle.
\end{equation}

The correlator in Eq.~\eqref{eq:O1u_keldysh} takes the form
$$\left\langle T_K\cdots\right\rangle=\frac{\gamma_{1\ua}\gamma_{1\da}\gamma_{2\ua}^*\gamma_{2\da}^*}{v_{\mathrm{P}\ua} v_{\mathrm{P}\da} v_{\mathrm{N}c}^3 v_{\mathrm{N}s}} e^{i(\varphi_\AB^\ua+\varphi_\AB^\da)} e^{ieV\left(t_1-t_2-t_3 + \frac{2l}{v_{\mathrm{N}c}}\right)}K(\vec t,l),$$
where $\vec t=(t_1,t_2,t_3)$ and $K(\vec t,l)= K_{\mathrm{P}}(\vec t,l)\,K_{\mathrm{N}c}(\vec t,l)\,K_{\mathrm{N}s}(\vec t,l)$. The three factors are the correlators of the electrons in the p region, the chargons, and the spinons, respectively,
\begin{equation}\label{eq:kernelK}
	\begin{aligned}
		K_{\mathrm{P}}(\vec t,l)=\;& \frac{v_{\mathrm{P}\ua} v_{\mathrm{P}\da}}{a^2} \left\langle T_K\,e^{i\phi_{\mathrm{P}\ua}(0,0^+)}e^{-i\phi_{\mathrm{P}\ua}(l,t_2^{\eta_2})}\right\rangle\left\langle T_K\,e^{i\phi_{\mathrm{P}\da}(0,t_1^{\eta_1})}e^{-i\phi_{\mathrm{P}\da}(l,t_3^{\eta_3})}\right\rangle \\
		=\;&		G_{-1}\left(\eta_2\left(-t_2+\frac{l}{v_{\mathrm{P}\ua}}\right)\right)
		G_{-1}\left(\chi_{\eta_1\eta_3}(t_1-t_3)
		\left(t_1-t_3+\frac{l}{v_{\mathrm{P}\da}}\right)\right),
		\\
		K_{\mathrm{N}c}(\vec t,l)=\;& \frac{v_{\mathrm{N}c}^3}{a^3} \left\langle T_K\,e^{-i\sqrt{\frac32}\phi_{\mathrm{N}c}(0,0^+)}e^{-i\sqrt{\frac32}\phi_{\mathrm{N}c}(0,t_1^{\eta_1})}e^{i\sqrt{\frac32}\phi_{\mathrm{N}c}(l,t_2^{\eta_2})}e^{i\sqrt{\frac32}\phi_{\mathrm{N}c}(l,t_3^{\eta_3})}\right\rangle \\
		=\;&
		G_{3/2}\left(-\eta_1 t_1\right)
		G_{3/2}\left(\chi_{\eta_2\eta_3}(t_2-t_3)(t_2-t_3)\right)
		G_{-3/2}\left(\eta_2\left(-t_2+\frac{l}{v_{\mathrm{N}c}}\right)\right)
		G_{-3/2}\left(\eta_3\left(-t_3+\frac{l}{v_{\mathrm{N}c}}\right)\right)
		\\
		&\times
		G_{-3/2}\left(\chi_{\eta_2\eta_1}(t_2-t_1)
		\left(t_2-t_1-\frac{l}{v_{\mathrm{N}c}}\right)\right)
		G_{-3/2}\left(\chi_{\eta_1\eta_3}(t_1-t_3)
		\left(t_1-t_3+\frac{l}{v_{\mathrm{N}c}}\right)\right),
		\\
		K_{\mathrm{N}s}(\vec t,l)=\;& \frac{v_{\mathrm{N}s}}{a} \left\langle T_K\,e^{-i\sqrt{\frac12}\phi_{\mathrm{N}s}(0,0^+)}e^{i\sqrt{\frac12}\phi_{\mathrm{N}s}(0,t_1^{\eta_1})}e^{i\sqrt{\frac12}\phi_{\mathrm{N}s}(l,t_2^{\eta_2})}e^{-i\sqrt{\frac12}\phi_{\mathrm{N}s}(l,t_3^{\eta_3})}\right\rangle \\
		=\;&
		G_{-1/2}\left(-\eta_1 t_1\right)
		G_{-1/2}\left(\chi_{\eta_2\eta_3}(t_2-t_3)(t_2-t_3)\right)
		G_{1/2}\left(\eta_3\left(-t_3-\frac{l}{v_{\mathrm{N}s}}\right)\right)
		G_{-1/2}\left(\eta_2\left(-t_2-\frac{l}{v_{\mathrm{N}s}}\right)\right)
		\\
		&\times
		G_{1/2}\left(\chi_{\eta_2\eta_1}(t_2-t_1)
		\left(t_2-t_1+\frac{l}{v_{\mathrm{N}s}}\right)\right)
		G_{-1/2}\left(\chi_{\eta_1\eta_3}(t_1-t_3)
		\left(t_1-t_3-\frac{l}{v_{\mathrm{N}s}}\right)\right).
	\end{aligned}
\end{equation}
Here $G_\alpha(t)=\left(\frac{i}{\pi T}\sinh[\pi T(t-ia)]\right)^\alpha$ and $\chi_{\eta_1\eta_2}(t)=\frac{\eta_1+\eta_2}{2}\sgn(t)-\frac{\eta_1-\eta_2}{2}$, and we used the bosonization $\psi_{\mathrm{P}\sigma}^\dagger=\frac{F_{\mathrm{P}\sigma}^\dagger}{\sqrt a}e^{i\phi_{\mathrm{P}\sigma}}$. Taking the derivative with respect to the voltage $V$ at $V=0$, we obtain
$$\frac{\partial\langle \O_{1\ua}(0)\rangle_{11}}{\partial V}=-e\frac{\gamma_{1\ua}\gamma_{1\da}\gamma_{2\ua}^*\gamma_{2\da}^*}{v_{\mathrm{P}\ua} v_{\mathrm{P}\da} v_{\mathrm{N}c}^3 v_{\mathrm{N}s}}e^{i(\varphi_\AB^\ua+\varphi_\AB^\da)} \sum_{\vec\eta}\eta_1\eta_2\eta_3\int_{t_1,t_2,t_3<0} \left(t_1-t_2-t_3 + \frac{2l}{v_{\mathrm{N}c}}\right) K(\vec t,l).$$
Similarly, Eq.~\eqref{eq:O2u_keldysh} yields
$$\frac{\partial\langle \O_{2\ua}^\dagger(0)\rangle_{11}}{\partial V}=e\frac{\gamma_{1\ua}\gamma_{1\da}\gamma_{2\ua}^*\gamma_{2\da}^*}{v_{\mathrm{P}\ua} v_{\mathrm{P}\da} v_{\mathrm{N}c}^3 v_{\mathrm{N}s}}e^{i(\varphi_\AB^\ua+\varphi_\AB^\da)}\sum_{\vec\eta}\eta_1\eta_2\eta_3\int_{t_1,t_2,t_3<0}\left(t_1-t_2-t_3-\frac{2l}{v_{\mathrm{N}c}}\right)K(\vec t,-l),$$
where $K(\vec t,-l)$ is obtained from Eq.~\eqref{eq:kernelK} by replacing $l\to -l$. Combining these results, the $(1,1)$ contribution to the differential conductance becomes
\begin{equation}\label{eq:11_conductance_general}
	\frac{\partial I_{11}}{\partial V}=-ie^2\frac{\gamma_{1\ua}\gamma_{1\da}\gamma_{2\ua}^*\gamma_{2\da}^*}{v_{\mathrm{P}\ua} v_{\mathrm{P}\da} v_{\mathrm{N}c}^3 v_{\mathrm{N}s}}e^{i(\varphi_\AB^\ua+\varphi_\AB^\da)}\left(W_{11}^\ua+W_{11}^\da\right)
\end{equation}
with
\begin{equation}\label{eq:W11_def}
	W_{11}^\ua=\sum_{\vec\eta}\eta_1\eta_2\eta_3\int_{t_1,t_2,t_3<0}\left(t_1-t_2-t_3+\frac{2l}{v_{\mathrm{N}c}}\right)K(\vec t,l)+(l\to -l).
\end{equation}
Here $W_{11}^\ua$ has the form of $F(l)+(l\to -l)\equiv F(l)+F(-l)$, and $W_{11}^\da$ is obtained from $W_{11}^\ua$ by exchanging $v_{\mathrm{P}\ua}\leftrightarrow v_{\mathrm{P}\da}$.

In the high-temperature limit, we approximate $K(\vec t,l)\sim e^{\pi T S(\vec t,l)}$, where
\begin{equation} \label{S11}
	\begin{aligned}
		S(\vec t,l)=\;&
		-\left|t_2-\frac{l}{v_{\mathrm{P}\ua}}\right|
		-\left|t_1-t_3+\frac{l}{v_{\mathrm{P}\da}}\right|
		+\left|t_1\right|+\left|t_2-t_3\right|\\
		&-\frac{3}{2}\left(
		\left|t_2-\frac{l}{v_{\mathrm{N}c}}\right|
		+\left|t_3-\frac{l}{v_{\mathrm{N}c}}\right|
		+\left|t_2-t_1-\frac{l}{v_{\mathrm{N}c}}\right|
		+\left|t_1-t_3+\frac{l}{v_{\mathrm{N}c}}\right|
		\right)\\
		&+\frac{1}{2}\left(
		\left|t_3+\frac{l}{v_{\mathrm{N}s}}\right|
		+\left|t_2-t_1+\frac{l}{v_{\mathrm{N}s}}\right|
		-\left|t_2+\frac{l}{v_{\mathrm{N}s}}\right|
		-\left|t_1-t_3-\frac{l}{v_{\mathrm{N}s}}\right|
		\right)
	\end{aligned}
\end{equation}
describes the temporal delays among the excitations between the two interfering subprocesses. 
For $v_{\mathrm{N}c}>v_{\mathrm{P}\ua},v_{\mathrm{P}\da}$, the overall maximum among the two functions $S(\vec t,\pm l)$ is attained in the $S(\vec t,-l)$ case at $t_1=0$ and $t_2=t_3=-l/v_{\mathrm{N}c}$, with the maximum value $S_* = -(\Delta t_\ua+\Delta t_\da)$. 
In this configuration, the spin-up and spin-down electrons tunnel simultaneously, at the upper beam splitter at time $-l/v_{\mathrm{N}c}$ in one subprocess and at the lower beam splitter at time $0$ in the other (Fig.~4d of the main text). The chargons then have the perfect overlap between the two interfering subprocesses. The up- and down-spinons generated at the beam splitter fuse into the vacuum channel, as discussed below. Only the electrons on the spin-up and spin-down channels in the p region exhibit temporal delays of $\Delta t_\ua$ and $\Delta t_\da$, respectively, which correspond to the first two terms of Eq.~\eqref{S11}; the other terms of Eq.~\eqref{S11} vanish. Accordingly, the dominant contribution to $W_{11}^\ua$ comes from the $K(\vec t,-l)$ term in Eq.~\eqref{eq:W11_def}, yielding $W_{11}^\ua \sim e^{\pi T S_*}$.

Around the dominant configuration, the up- and down-spinons generated at each beam splitter are temporally separated by only $O(1/T)$, much shorter than the delay $\Delta t_s$ between the spinons generated at the upper beam splitter and those generated at the lower one. By spinon fusion, the spinon correlator $K_{\mathrm{N}s}(\vec t,-l)$ factorizes into those of the up- and down-spinons at each beam splitter,
\begin{equation}\label{eq:spinon_fusion}
	\begin{aligned}
		K_{\mathrm{N}s}(\vec t,-l) &= \frac{v_{\mathrm{N}s}}{a} \left\langle T_K\,e^{i\sqrt{\frac12}\phi_{\mathrm{N}s}(l,0^+)}e^{-i\sqrt{\frac12}\phi_{\mathrm{N}s}(l,t_1^{\eta_1})}e^{-i\sqrt{\frac12}\phi_{\mathrm{N}s}(0,t_2^{\eta_2})}e^{i\sqrt{\frac12}\phi_{\mathrm{N}s}(0,t_3^{\eta_3})}\right\rangle \\
		&\approx \frac{v_{\mathrm{N}s}}{a} \left\langle T_K\,e^{i\sqrt{\frac12}\phi_{\mathrm{N}s}(l,0^+)}e^{-i\sqrt{\frac12}\phi_{\mathrm{N}s}(l,t_1^{\eta_1})} \right\rangle \left\langle T_K\, e^{-i\sqrt{\frac12}\phi_{\mathrm{N}s}(0,t_2^{\eta_2})}e^{i\sqrt{\frac12}\phi_{\mathrm{N}s}(0,t_3^{\eta_3})}\right\rangle \\
		&= G_{-1/2}\left(-\eta_1 t_1\right) G_{-1/2}\left(\chi_{\eta_2\eta_3}(t_2-t_3)(t_2-t_3)\right),
	\end{aligned}
\end{equation}
up to corrections of $O(e^{-2\pi T\Delta t_s})$. This factorization is a manifestation of the fusion rule $1/2\times1/2=0$ of the SU(2)$_1$ conformal field theory, where the operator product expansion of the up- and down-spinons involves the identity operator $\mathbbm 1$ and its descendants,
\begin{equation}\label{eq:spinon_OPE}
	e^{i\sqrt{\frac12}\phi_{\mathrm{N}s}(x,t)}e^{-i\sqrt{\frac12}\phi_{\mathrm{N}s}(x,t')}\propto [a+i(t-t')]^{-1/2}\,\mathbbm 1+\cdots.
\end{equation}
As a result, $K_{\mathrm{N}s}(\vec t,-l)$ does not depend on the arm length $l$, and the spin mode causes no exponential thermal suppression of the interference by $\Delta t_s$.

To extract the prefactor, we consider fluctuations around this point by introducing $u_1=\pi T t_1$ and $u_{2,3}=\pi T\left(t_{2,3}+\frac{l}{v_{\mathrm{N}c}}\right)$. Performing a calculation similar to that leading to Eq.~\eqref{10result}, we obtain
\begin{equation}\label{eq:W11_prefactor}
	W_{11}^\ua\approx C T^2 e^{-\pi T(\Delta t_\ua+\Delta t_\da)},
\end{equation}
where the dimensionless constant $C$ is
$$C=8\pi^2\int_{u_1<0,u_2,u_3\in\mathbb R}\sum_{\eta_2,\eta_3\in\{\pm 1\}}
\frac{(u_1-u_2-u_3)e^{u_1-u_2-u_3}\tilde G_1(u_1)\tilde G_1(\chi_{\eta_2\eta_3}(u_2-u_3)(u_2-u_3))}
{\tilde G_{3/2}(-\eta_2u_2)\tilde G_{3/2}(-\eta_3u_3)\tilde G_{3/2}(\eta_2(u_1-u_2))\tilde G_{3/2}(\eta_3(u_1-u_3))}$$
with $\tilde G_\alpha(u)=[i\sinh(u-ia)]^\alpha$. Numerical integration gives $C\approx 7938i$.
Finally, substituting Eq.~\eqref{eq:W11_prefactor} into Eq.~\eqref{eq:11_conductance_general}, we obtain Eq.~\eqref{eq:I11thermal} of the main text,
\begin{equation}\label{eq:11_final}
	\frac{\partial I_{11}}{\partial V}\approx -ie^2\frac{\gamma_{1\ua}\gamma_{1\da}\gamma_{2\ua}^*\gamma_{2\da}^*}{v_{\mathrm{P}\ua} v_{\mathrm{P}\da} v_{\mathrm{N}c}^3 v_{\mathrm{N}s}}e^{i(\varphi_\AB^\ua+\varphi_\AB^\da)}(2C)T^2 e^{-\pi T(\Delta t_\ua+\Delta t_\da)}.
\end{equation}
This result is confirmed by direct numerical integration of Eq.~\eqref{eq:W11_def}.

\supplementarynote{7}{Comparison between experiments and theories}
\label{SN7}

\begin{figure}[h]
	\centering
	\includegraphics{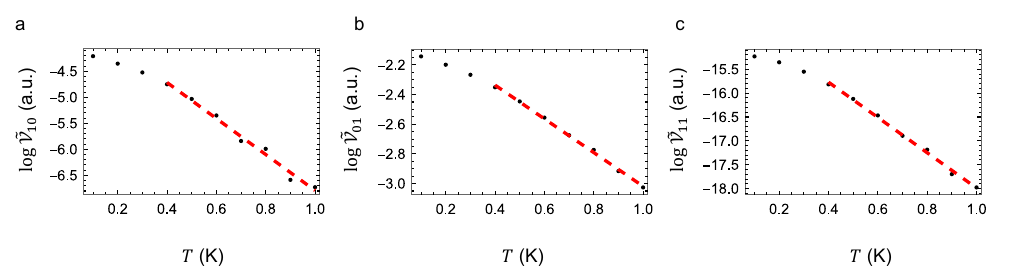}
	\caption{Temperature dependence of the generalized visibilities $\Vis_{nm}$ (in arbitrary units) of (a) interference frequencies $(n,m)=(1,0)$, (b) $(n,m)=(0,1)$, and (c) $(n,m)=(1,1)$. Dots represent experimental data of $\log\Vis_{nm}$. At high temperature, the visibilities approximately follow exponential decay. The dashed lines are linear fits to $\log\Vis_{nm}$, performed after excluding the three lowest-temperature data points. From the fit, the decay rates $B_{nm}$ of the visibilities $\Vis_{nm} \sim \exp\left(-B_{nm}\frac{\pi k_BT}{\hbar}\right)$ are extracted.
	}
	\label{visibilityfit}	
\end{figure}

The interference visibilities in Eqs.~\eqref{eq:I10thermal}, \eqref{eq:I01thermal}, \eqref{eq:I11thermal} of the main text depend on the time delays $\Delta t_\sigma=\frac{l}{v_{\mathrm{P}\sigma}}-\frac{l}{v_{\mathrm{N}c}}$ and $\Delta t_s=\frac{l}{v_{\mathrm{N}s}}+\frac{l}{v_{\mathrm{N}c}}$.
We extract the time delays from the experimental data on the temperature dependence of the visibility for the $(n,m) = (1,0)$, $(0,1)$, and $(1,1)$ interference components. Here $v_{\mathrm{P}\sigma}$ denotes the velocity of the spin-$\sigma$ mode in the p region, while $v_{\mathrm{N}c}$ and $v_{\mathrm{N}s}$ are the charge and spin mode velocities in the n region.

For the three interference components, the temperature-dependent visibility at high temperature has the form $|\partial I_{nm}/\partial V| \sim T^{n+m} \exp\left(-B_{nm}\frac{\pi k_BT}{\hbar}\right)$, as shown in Eqs.~\eqref{eq:I10thermal}, \eqref{eq:I01thermal}, \eqref{eq:I11thermal}. 
The power-law prefactor is not quantitatively reproduced in our experiment, hence we focus on the exponential decay characterized by $B_{nm}$. To this aim, we introduce the ``generalized'' visibility $\Vis_{nm}=\frac{|\partial I_{nm}/\partial V|}{(\partial I_{00}/\partial V)^{n+m}}$.
Compared with the visibility $\mathcal{V}_{nm}$ in Fig.~3e of the main text, the denominator $(\partial I_{00}/\partial V)^{n+m}$ is introduced to cancel the power-law part of $|\partial I_{nm}/\partial V|$, since $\partial I_{00}/\partial V\propto T$.
According to Eqs.~\eqref{eq:I10thermal}, \eqref{eq:I01thermal}, \eqref{eq:I11thermal}, the decay rates $B_{nm}$ are given by
\begin{equation} \label{decayrates}
	B_{10}=\Delta t_\ua+\frac{1}{2}\Delta t_s,\quad B_{01}=\Delta t_\da+\frac{1}{2}\Delta t_s,\quad B_{11}=\Delta t_\ua+\Delta t_\da
\end{equation}
in the regime where $v_{\mathrm{N}c}>v_{\mathrm{P}\ua},v_{\mathrm{P}\da}$ and the intra-channel interactions are negligible.

Indeed, the generalized visibilities measured in our experiments approximately follow exponential decay at high temperature, as shown in Fig.~\ref{visibilityfit}.
We fit $\log\Vis_{nm}$ linearly as a function of temperature after excluding the three lowest-temperature data points. From the fitting, we estimate $B_{10}=\SI{8.41}{ps}$, $B_{01}=\SI{2.76}{ps}$, $B_{11}=\SI{8.98}{ps}$.
Solving the linear equations in Eq.~\eqref{decayrates}, we obtain
$$\Delta t_\ua=\SI{7.31}{ps},\quad \Delta t_\da=\SI{1.67}{ps},\quad \Delta t_s=\SI{2.18}{ps}.$$

The extracted time delays are on the picosecond scale, as expected from the typical edge velocity $v\sim\SI{1e5}{m/s}$ and the interferometer arm length $l=\SI{700}{nm}$ in our experiment.
They are also consistent with the velocity ordering $v_{\mathrm{N}c}>v_{\mathrm{P}\ua},v_{\mathrm{P}\da}$ (equivalent to $\Delta t_\ua, \Delta t_\da >0$) assumed in the derivation of Eqs.~\eqref{eq:I10thermal}, \eqref{eq:I01thermal}, \eqref{eq:I11thermal}. They suggest $\Delta t_s<\Delta t_\ua$, interestingly implying that the spin mode in the $\nu_{\mathrm{N}}=2/3$ region is faster than the spin-up electrons of the $\nu_{\mathrm{P}} = -2$ region.


This extraction is rendered inconclusive by two possibilities: first, that our model considers only delta function-like interchannel interactions,  neglecting other possible interactions present in the experiment; 
and second,
that our experiments may not be in the regime of the assumed velocity ordering $v_{\mathrm{N}c}>v_{\mathrm{P}\ua},v_{\mathrm{P}\da}$.
Note that we do not derive the full conductance for the $(0,2)$ component.

\supplementarynote{8}{Effects of SU(2) symmetry breaking}
\label{SN8}

In the main text, we focused on the case where the SU(2) symmetry is preserved along the spin-unpolarized $\nu=2/3$ edge.
Here we consider the situation that this symmetry is slightly broken, and show that even in this more general regime, the electron fractionalization still occurs and the $(n,m)$ interference components remain present.

Within the $K$-matrix formalism~\cite{wen1992Kmatrix}, the unpolarized $\nu=2/3$ edge is described by two boson fields $\phi_\ua$ and $\phi_\da$. They satisfy the commutation relation $[\phi_i(x_1), \phi_j(x_2)] = i\pi K^{-1}_{ij} \sgn(x_1-x_2)$, where the $K$ matrix is
$K = \begin{pmatrix}
	1&2\\2&1
\end{pmatrix}$
in the $\{\phi_\ua, \phi_\da\}$ basis. The kinetic Hamiltonian is
$$H_0 = \frac{1}{4\pi} \sum_{i,j\in\{\ua,\da\}} \int dx \: V_{ij} \,\partial_x \phi_i \, \partial_x\phi_j$$
with the interaction matrix
$$V = \begin{pmatrix}
	v_\ua & v_{\ua\da} \\ v_{\ua\da} & v_\da
\end{pmatrix}.$$
The SU(2)-symmetric case corresponds to $v_\ua = v_\da$, whereas we allow more general cases $v_\ua \neq v_\da$ here.

To obtain the eigenmodes, we perform the transformation
$$\begin{pmatrix} \phi_\ua \\ \phi_\da \end{pmatrix} = R \begin{pmatrix} \phi_d \\ \phi_u \end{pmatrix}, \quad R = \begin{pmatrix}
	\sqrt{\frac16} \cosh \theta + \sqrt{\frac12} \sinh \theta & - \sqrt{\frac12} \cosh \theta - \sqrt{\frac16} \sinh \theta \\
	\sqrt{\frac16} \cosh \theta - \sqrt{\frac12} \sinh \theta & \sqrt{\frac12} \cosh \theta - \sqrt{\frac16} \sinh \theta
\end{pmatrix},$$
where $\theta$ satisfies $\tanh 2 \theta = \frac{\sqrt 3(v_\da-v_\ua)}{2(v_\ua+v_\da-v_{\ua\da})}$. The Hamiltonian then takes the diagonal form
$$H_0 = \int dx \left[ \frac{v_d}{4\pi}(\partial_x\phi_d)^2 + \frac{v_u}{4\pi}(\partial_x\phi_u)^2 \right].$$
Here the downstream mode $\phi_d$ and upstream mode $\phi_u$ satisfy $[\phi_d(x_1), \phi_d(x_2)] = i\pi \sgn(x_1-x_2)$, $[\phi_u(x_1), \phi_u(x_2)] = -i\pi \sgn(x_1-x_2)$, and $[\phi_d(x_1), \phi_u(x_2)] = 0$. Their velocities are
\begin{equation} \label{modevelocity}
	v_{d,u} = \frac16 \left[ \sqrt{4(v_\ua+v_\da-v_{\ua\da})^2 - 3(v_\ua-v_\da)^2} \mp (v_\ua+v_\da-4v_{\ua\da}) \right].
\end{equation}
In the SU(2)-symmetric limit $v_\ua = v_\da$, one has $\theta=0$, and the downstream and upstream modes reduce to the charge and spin modes, $\phi_c = \sqrt{\frac32}(\phi_\ua+\phi_\da)$ and $\phi_s = -\sqrt{\frac12}(\phi_\ua-\phi_\da)$, respectively. In this case, the Hamiltonian takes the same form as Eq.~\eqref{eq:H0} of the main text, with the corresponding velocities $v_c = \frac13(v_\ua + v_{\ua\da})$ and $v_s = v_\ua - v_{\ua\da}$. Note that the charge (spin) mode becomes faster (slower) by the repulsive interaction $v_{\ua\da}>0$.

In the $K$-matrix formalism~\cite{wen1992Kmatrix}, quasiparticle operators take the form $e^{i(n_\ua \phi_\ua + n_\da \phi_\da)}$, where $\mathbf n=(n_\ua,n_\da)^T$ is an integer vector. Such a quasiparticle carries charge $q=-e\,\mathbf t^T K^{-1}\mathbf n$, where $\mathbf t=(1,1)^T$ is the charge vector. The mutual statistical angle between two quasiparticles labeled by $\mathbf n$ and $\mathbf n'$ is given by $\vartheta=2\pi\,\mathbf n^T K^{-1}\mathbf n'$. When $\mb n$ is a column of $K$, i.e., $\mathbf n=K\mathbf e_i$ with the $i$-th standard unit vector $\mb e_i$,  the corresponding quasiparticle carries charge $-e$ and has trivial braiding statistics with all other quasiparticles. It can therefore be identified as an electron. In our setup, the spin-up and spin-down electron creation operators are thus given by $\psi_\ua^\dagger \propto e^{i\phi_\ua}e^{2i\phi_\da}$ and $\psi_\da^\dagger \propto e^{2i\phi_\ua}e^{i\phi_\da}$, respectively. 
In terms of the eigenmodes, they take the form $\psi_\sigma^\dagger \propto e^{i\lambda_{d\sigma} \phi_d} e^{i\lambda_{u\sigma} \phi_u}$, where
$$\lambda_{d\sigma} = \sqrt{\frac32} \cosh \theta -\sigma \sqrt{\frac12} \sinh\theta, \quad \lambda_{u\sigma} = -\sqrt{\frac32} \sinh\theta +\sigma \sqrt{\frac12} \cosh\theta$$
with $\sigma=+1$ for $\ua$ and $\sigma=-1$ for $\da$. This form makes the electron fractionalization explicit: an injected electron is decomposed into downstream and upstream components associated with the two eigenmodes. The corresponding fractionalized charges are $q_{d\sigma} = -e \sqrt{\frac23} \lambda_{d\sigma} \cosh\theta$ and $q_{u\sigma} = -e \sqrt{\frac23} \lambda_{u\sigma}\sinh\theta$, respectively, which satisfy $q_{d\sigma} + q_{u\sigma}=-e$. In the SU(2)-symmetric case $\theta=0$, the electron operators become $\psi_\sigma^\dagger \propto e^{i\sqrt{\frac32}\phi_c} e^{i\sigma\sqrt{\frac12} \phi_s}$ as in the main text, and we have $q_{d\sigma} = -e$, $q_{u\sigma}=0$.

\begin{figure}
	\centering
	\includegraphics{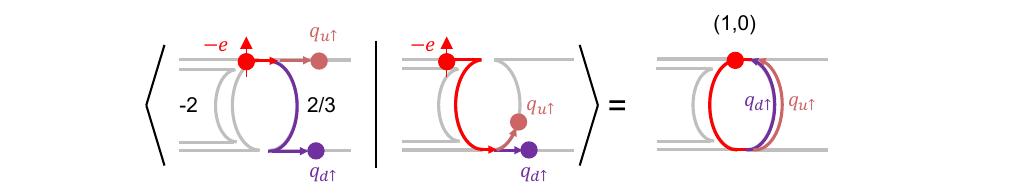}
	\caption{Schematic of the $(1,0)$ interference process in the spin-unpolarized regime with broken SU(2) symmetry. A spin-up electron injected from the $\nu_{\mathrm{P}}=-2$ region fractionalizes at the beam splitter into downstream and upstream excitations carrying charges $q_{d\ua}$ and $q_{u\ua}$, respectively. Their combined interference nevertheless produces the same AB phase as electron winding, since $q_{d\ua}+q_{u\ua}=-e$. The SU(2)-symmetric case in Fig.~4c of the main text corresponds to $q_{u\ua}=0$.
	}
	\label{su2breaking}	
\end{figure}

Even when the SU(2) symmetry is broken, we find that the $(n,m)$ interferences discussed in the main text still occur, and that their AB oscillation frequencies still correspond to the winding of electron charge. As an example, Fig.~\ref{su2breaking} illustrates the $(1,0)$ interference process for the broken-symmetry case. In the $\nu_{\mathrm{N}}=2/3$ region, the downstream and upstream components with charges $q_{d\ua}$ and $q_{u\ua}$ together produce the AB phase associated with electron winding. It should be noted that the expression of the interference conductance is modified from the symmetric case. For example, the conductance in Eq.~\eqref{eq:I10thermal} of the main text becomes modified into $\frac{\partial I_{10}}{\partial V} \propto T^{2h_\ua-1} e^{-\pi T ( \Delta t_\ua + 2h_{u\ua} \Delta t_u)}$, where $h_\ua = (\lambda_{d\ua}^2 + \lambda_{u\ua}^2)/2$ is the scaling dimension of $\psi_\ua^\dagger$, $h_{u\ua} = \lambda_{u\ua}^2/2$ is the scaling dimension of its upstream component, and $\Delta t_\ua = \frac{l}{v_{\mathrm{P}\ua}} - \frac{l}{v_{\mathrm{N}d}}$, $\Delta t_u = \frac{l}{v_{\mathrm{N}u}} + \frac{l}{v_{\mathrm{N}d}}$. Here $v_{\mathrm{N}d}$ and $v_{\mathrm{N}u}$ are the mode velocities in the n region in Eq.~\eqref{modevelocity}. In deriving this expression, we assumed $v_{\mathrm{N}d} > v_{\mathrm{P}\ua}$ and considered the high-temperature regime $k_BT \gg \frac{\hbar}{\Delta t_\ua}, \frac{\hbar}{\Delta t_u}$.

\supplementarynote{9}{Peierls substitution}
\label{SN9}

In the Methods of the main text, the applied voltage is incorporated into the tunneling Hamiltonian as a Peierls phase. Here, we derive the Peierls phase in the spin-unpolarized regime [see Eq.~\eqref{eq:HT} of the main text]; a similar derivation applies to Eq.~\eqref{HTpl} in the spin-polarized regime.

We denote the Hamiltonian before the Peierls substitution with a tilde, writing it as $\tilde H = H_0 + \tilde H_V + \tilde H_T$. Here, $H_0$ is the kinetic Hamiltonian in Eq.~\eqref{eq:H0} (see the main text), $\tilde H_V = V\int_{-\infty}^d dx\: \rho_{\mathrm{N}}(x)$ is the voltage term, and $\tilde H_T$ is the tunneling Hamiltonian before the Peierls substitution. The voltage $V$ is applied to the n region for $x<d\,(<0)$, i.e., before the two tunneling points at $x=0,l$, and $\rho_{\mathrm{N}}(x) = \frac{e}{2\pi} [\partial_x\phi_\ua(x) + \partial_x\phi_\da(x)]=\frac{e}{\sqrt 6\pi} \partial_x\phi_{\mathrm{N}c}$ describes the electric charge density. Note that $\rho_{\mathrm{N}}(x)$ involves only the charge mode $\phi_{\mathrm{N}c}$; the neutral spin mode $\phi_{\mathrm{N}s}$ does not enter. 

The tunneling Hamiltonian is written as
$\tilde H_T = \sum_{\sigma\in\{\ua,\da\}} [\tilde{\mathcal{O}}_{1\sigma} + \tilde{\mathcal{O}}_{2\sigma}] + \textrm{h.c.}$,
where $\tilde{\mathcal{O}}_{1\sigma} = \gamma_{1\sigma} e^{i\varphi_\AB^\sigma} \psi_{\mathrm{P}\sigma}^\dagger(0)\psi_{\mathrm{N}\sigma}(0)$ and $\tilde{\mathcal{O}}_{2\sigma} = \gamma_{2\sigma}  \psi_{\mathrm{P}\sigma}^\dagger(l)\psi_{\mathrm{N}\sigma}(l)$ do not yet include the Peierls phase.

To implement the Peierls substitution, we use the following fact: Under a unitary transformation $\Psi(t) = U(t) \tilde\Psi(t)$, a state $\tilde\Psi(t)$ satisfying the time-dependent Schr\"odinger equation $i\partial_t \tilde\Psi = \tilde H(t) \tilde\Psi$ obeys the transformed dynamics $i\partial_t \Psi = H(t) \Psi$. The transformed Hamiltonian is given by $H(t) = U(t) \tilde H(t) U^\dagger(t) + i [\partial_t U(t)] \: U^\dagger(t)$. Choosing $U(t) = e^{-iH_0(t-t_0)} e^{i(H_0+\tilde H_V)(t-t_0)}$ with $t_0 =-\infty$, we remove $\tilde H_V$ in the transformed Hamiltonian as
$$H(t) = H_0 + U(t) \tilde H_T U^\dagger(t).$$
For further computation, it is sufficient to determine how $\phi_{\mathrm{N}c}$ transforms under $U(t)$, 
since $\tilde H_V$ couples only to the charge mode in the n region. For the purpose, we introduce
$$\Phi_{\mathrm{N}c}(x,t) := e^{i(H_0+\tilde H_V)(t-t_0)} \phi_{\mathrm{N}c}(x) e^{-i(H_0+\tilde H_V)(t-t_0)},$$
so that $U(t) \phi_{\mathrm{N}c}(x) U^\dagger(t) = e^{-iH_0(t-t_0)} \Phi_{\mathrm{N}c}(x,t) e^{iH_0(t-t_0)}$. The field $\Phi_{\mathrm{N}c}(x,t)$ satisfies the equation of motion
$$(\partial_t + v_{\mathrm{N}c} \partial_x) \Phi_{\mathrm{N}c}(x,t) = -\sqrt{\frac23} eV \Theta(d-x),$$
where $\Theta$ is the Heaviside function; the voltage acts as a source term on $x<d$. Its solution is
$$\Phi_{\mathrm{N}c}(x,t)=  \phi_{\mathrm{N}c}(x-v_{\mathrm{N}c}(t-t_0)) -\sqrt{\frac23} eV \int_{t_0}^t dt' \: \Theta(d-x+v_{\mathrm{N}c}(t-t')).$$
The integral is interpreted as the time interval during which a chiral excitation arriving at $(x,t)$ has spent in the voltage-biased region when traced backward in time. For $x>d$, the integral simplifies to $t-t_0 - \frac{x-d}{v_{\mathrm{N}c}}$, and $\phi_{\mathrm{N}c}$ satisfies
$$U(t) \phi_{\mathrm{N}c}(x) U^\dagger(t) = \phi_{\mathrm{N}c}(x) -  \sqrt{\frac23} eV \left(t-t_0 - \frac{x-d}{v_{\mathrm{N}c}}\right).$$
As a result, the electron operator in the n region $\psi_{\mathrm{N}\sigma}^\dagger(x) \propto e^{i\sqrt{\frac32}\phi_{\mathrm{N}c}(x)} e^{i\sigma\sqrt{\frac12}\phi_{\mathrm{N}s}(x)}$ acquires an additional phase factor under the transformation,
$U(t) \psi_{\mathrm{N}\sigma}^\dagger(x) U^\dagger(t) = e^{-ieV\left(t-t_0 - \frac{x-d}{v_{\mathrm{N}c}}\right)}\psi_{\mathrm{N}\sigma}^\dagger(x)$.
Then the tunneling Hamiltonian transforms,
$$U(t) \tilde H_T U^\dagger(t) = e^{-ieV\left(t_0-\frac{d}{v_{\mathrm{N}c}}\right)}\sum_{\sigma\in\{\ua,\da\}} \left[ e^{ieVt}\tilde{\mathcal{O}}_{1\sigma} + e^{ieV\left(t-\frac{l}{v_{\mathrm{N}c}}\right)}\tilde{\mathcal{O}}_{2\sigma} \right] + \textrm{h.c.}$$
As the overall phase factor $e^{-ieV\left(t_0-\frac{d}{v_{\mathrm{N}c}}\right)}$ is irrelevant and can be dropped, this leads to the tunneling Hamiltonian $H_T$ in Eq.~\eqref{eq:HT} of the main text.

%

\end{document}